\documentclass[12pt,a4paper]{article}

\usepackage{epsfig}
\usepackage{amssymb}
\usepackage{xcolor}
\usepackage[centerlast,footnotesize]{caption2}
\usepackage{float}

\newcommand{\lapprox}{\raisebox{-0.5ex}{$\
\stackrel{\textstyle<}{\textstyle\sim}\ $}}
\newcommand{\gapprox}{\raisebox{-0.5ex}{$\
\stackrel{\textstyle>}{\textstyle\sim}\ $}}

\usepackage{graphicx}
\newcommand{\One}{1\kern-4.5pt1}

\newcommand{\be}{\begin{equation}}
\newcommand{\ee}{\end{equation}}

\def\lesim{${\lower 2pt\hbox{$\scriptstyle
<$}\atop\raise 4pt\hbox{$\scriptstyle\sim$}}$} 
\def\grsim{${\lower2pt\hbox{$\scriptstyle >$} \atop\raise4pt\hbox 
{$\scriptstyle\sim$}}$} 

\begin{document}
\begin{center}
\begin{flushright}
November 2018
\end{flushright}
\vskip 10mm
{\LARGE
On the Critical Flavor Number \\ \vskip 0.3cm in the 2+1$d$ Thirring Model 
}
\vskip 0.3 cm
{\bf Simon Hands}
\vskip 0.3 cm
{\em Department of Physics, College of Science, Swansea University,\\
Singleton Park, Swansea SA2 8PP, United Kingdom.}
\end{center}

\vskip 1.3 cm
\noindent
{\bf Abstract:} 
The Thirring model in 2+1 spacetime dimensions, in which $N$ flavors of
relativistic fermion
interact via a contact interaction between conserved fermion currents, is studied using lattice field
theory simulations employing domain wall fermions, which furnish the correct U($2N$)
global symmetry in the limit that the wall separation $L_s\to\infty$.
Attention is focussed on the issue of spontaneous symmetry breakdown via a 
non-vanishing fermion bilinear condensate $\langle\bar\psi\psi\rangle\not=0$.
Results from quenched simulations are presented demonstrating that 
a non-zero condensate does indeed form over a range of couplings, provided 
simulation results are first extrapolated to the $L_s\to\infty$ limit.
Next, results from simulations with $N=1$ using an RHMC algorithm 
demonstrate that U(2) symmetry is unbroken at weak coupling but
plausibly broken at
strong coupling. Correlators of mesons with spin zero are consistent with the
Goldstone spectrum expected from U(2)$\to$U(1)$\otimes$U(1). We infer the
existence of a symmetry-breaking phase transition at some finite coupling, and 
combine this with previous simulation results to deduce that the critical number of
flavors for the existence of a quantum critical point in the Thirring model
satisfies $0<N_c<2$, with strong evidence that in fact
$N_c>1$.

\vspace{0.5cm}


\newpage

\section{Introduction}
\label{sec:intro}
While the study of theories of relativistic fermions moving in the plane, ie. in 2+1 spacetime
dimensions, has received a fillip over the past decade as a result of
developments in the condensed matter physics of layered systems, the
underlying quantum field theories continue
to be of considerable theoretical interest in their own right. This paper
concerns the $d=2+1$ Thirring model, describing $N$ flavors of interacting
fermion, whose Lagrangian density reads
\begin{equation}
{\cal
L}=\bar\psi_i(\partial\!\!\!/\,+m)\psi_i+{g^2\over{2N}}(\bar\psi_i\gamma_\mu\psi_i)^2;\;\;
i=1,\ldots,N.
\label{eq:L}
\end{equation}
Here $\mu=0,1,2$, and a key feature is that the fields $\psi$,$\bar\psi$ lie in 
{\em reducible\/} spinor representations, so that Dirac matrices
$\gamma_\mu$ are $4\times4$. This enables the definition of a mass term
$m\bar\psi\psi$ which is parity-invariant, where it is convenient to define a
discrete parity transformation in terms of inversion of all three spacetime
axes.

One reason the Thirring model is interesting is its unexpected
renormalisability; while a 
naive expansion in powers of the coupling $g^2$ is no longer power-counting
renormalisable once $d>2$, if a regularisation is chosen such that current 
conservation $\partial_\mu(\bar\psi\gamma_\mu\psi)=0$ is respected then an
expansion in powers of $1/N$ is exactly renormalisable over a continuum of
dimensionality $d\in(2,4)$, and moreover $g^2$ turns out to be
marginal~\cite{Parisi:1975im} -- \cite{Hands:1994kb}.
More interesting still is the possibility that the true ground state
has a bilinear condensate $\langle\bar\psi\psi\rangle\not=0$;
fermions propagating through such a vacuum incur a dynamically-generated
mass. This phenomenon can be described in terms of spontaneous breaking of a 
global U(2$N$) symmetry. This follows since with two $4\times4$ Dirac matrices anticommuting with
the kinetic operator in (\ref{eq:L}), for $m=0$ the following rotations leave
${\cal L}$ invariant:
\begin{eqnarray}
\psi\mapsto e^{i\alpha}\psi,\;\bar\psi\mapsto\bar\psi e^{-i\alpha};\;\;\; 
\psi\mapsto e^{\alpha\gamma_3\gamma_5}\psi,\;\bar\psi\mapsto\bar\psi
e^{-\alpha\gamma_3\gamma_5};\label{eq:mass}\\
\psi\mapsto e^{i\alpha\gamma_3}\psi,\;\bar\psi\mapsto\bar\psi
e^{i\alpha\gamma_3};\;\;\; \;\;\;
\psi\mapsto e^{i\alpha\gamma_5}\psi,\;\bar\psi\mapsto\bar\psi
e^{i\alpha\gamma_5}.\label{eq:massless}
\end{eqnarray}
Once $m\not=0$ only (\ref{eq:mass}) remain as symmetries; ie. dynamical fermion mass
generation corresponds to a breaking pattern U(2$N)\to$U($N)\otimes$U($N$).

The question  of whether bilinear condensation takes place is inherently
non-per\-turbative,
and has been first studied using truncated Schwinger-Dyson
equations~\cite{Gomes:1990ed,Itoh:1994cr,Sugiura:1996xk}. While details are
somewhat scheme-dependent, the picture emerging is that symmetry breaking is
possible for sufficiently large $g^2$ and sufficiently small $N$, and indeed
there exists a critical flavor number $N_c$ such that no symmetry breaking
occurs at any coupling for $N>N_c$. More recently the Functional
Renormalisation Group (FRG) has also been applied~\cite{Gies:2010st}.
The main focus of this work has been to establish the existence of
UV-stable RG fixed points $g_*^2(N)$, such that an
interacting field theory exists at all scales in the limit $g^2\to g_*^2$.
On the assumption that the symmetry-breaking transition is second-order, 
it seems reasonable to identify this quantum critical point (QCP) with the critical
$g_c^2(N)$, which exists for $N<N_c$. The identification of $N_c$ is thus an important ingredient in the
search for novel QCPs.

It is natural to apply lattice field theory methods to the problem, and
indeed this has been tried by several groups over the
years~\cite{DelDebbio:1997dv}  -- \cite{
Chandrasekharan:2011mn}. A common feature of all these approaches is the use of
staggered lattice fermions in 2+1$d$. The conclusion of \cite{Christofi:2007ye},
employing  simulation studies in the effective strong-coupling limit, 
is that $N_c=6.6(1)$, and that the critical exponent $\delta$ characterising
the response of the order parameter to an explicit symmetry-breaking mass at
criticality has value $\delta(N_c)\approx7$. Away from the strong-coupling
limit the value of $\delta$ is found to be rather sensitive to $N$. This is to be compared with the
Schwinger-Dyson predictions $N_c\simeq4.32$, $\delta(N_c)=1$~\cite{Itoh:1994cr}.
In summary, the staggered Thirring model exhibits a non-trivial phase
diagram in the $(N,g^2)$ plane with some interesting features. 

At the level of the lattice action, massless staggered fermions in 2+1$d$ have a manifest
U($N_{\rm stag})\otimes$U($N_{\rm stag}$) global symmetry, distinct from the U($2N$) of
eqn.~(\ref{eq:L}), which is broken to U($N_{\rm stag}$) by a fermion mass.
In a weak-coupling continuum limit, $N_{\rm stag}$ staggered
fermions are known to describe $N=2N_{\rm stag}$ continuum flavors, with an
eventual recovery of
U($2N$) at long wavelengths~\cite{Burden:1986by}. Near a QCP, however, the story may be different.
Indeed, a study of the $N=2$ staggered model using a fermion bag
algorithm~\cite{Chandrasekharan:2011mn}, 
which permits simulations directly
in the massless limit, found critical exponents compatible with those of the
Gross-Neveu model~\cite{Chandrasekharan:2013aya}, an unxepected result since on
the face of it the
two models have distinct Lagrangians, different symmetries, and completely different 
$1/N$ expansions, with symmetry breaking due to bilinear condensation
predicted in the large-$N$ limit in the GN case, and 
expected to persist  for all $N$. Instead, the results of
\cite{Chandrasekharan:2011mn,Chandrasekharan:2013aya} imply the two models lie
in the same RG basin of attraction.  Indeed, when written purely in terms of
four-point interactions between staggered fermion fields spread over the vertices of
elementary cubes, the only difference between Thirring and GN is an extra
body-diagonal coupling in the latter case~\cite{Chandrasekharan:2013aya}.

The mismatch between theoretical expectation and results from the staggered
Thir\-ring model has motivated us to consider alternative lattice fermion
formulations with the potential to capture the resquisite symmetries more
faithfully. Specifically, we have developed both analytical and numerical 
insight into how U(2$N$)  symmetry
is manifested in 2+1$d$ using Domain Wall
Fermions (DWF)~\cite{Hands:2015qha,Hands:2015dyp}, which will be reviewed in the next
section. Next, in Ref.~\cite{Hands:2016foa} we applied DWF in exploratory
simulations of both GN and Thirring models with $N=2$. The main conclusions of
that work were that in the GN model there appears to be no obstruction to studying symmetry
breaking via bilinear condensation and the resulting QCP; 
good qualitative
agreement was found with the analytical expectations of the large-$N$ approach.
However, no evidence was found for symmetry breaking in the Thirring model with
fixed $L_s=16$,
implying $N_c<2$ in contradiction to the staggered fermion model. 
Meanwhile, the Jena group has applied another U($2N$)-invariant formulation, the
SLAC fermion, and found no symmetry breaking in the Thirring model all the way down
to $N=1$~\cite{Wellegehausen:2017goy}. We also note in this context the recent
reported mismatch between DWF and staggered fermion results near a conformal
fixed point in 3+1$d$ non-abelian gauge theory, which potentially springs from
the same failure of the staggered action to capture the correct symmetries away
from weak coupling~\cite{Hasenfratz:2017mdh}.

This paper extends the study of spontaneous symmetry breaking in
the Thirring model with DWF to $N=1$. The formulation is
reviewed below in Sec.~\ref{sec:formulation}. Of the two possible ways to
introduce the Thirring interaction discussed in \cite{Hands:2016foa}, we will
focus almost exclusively on the variant in which the auxiliary vector field is
located uniformly throughout the bulk, ie. stressing its resemblence to an
abelian gauge potential. Simulating just a single DWF flavor requires the RHMC
algorithm, as argued in Appendix~\ref{app:A}; details of the implementation are
also given in Sec.~\ref{sec:formulation}, along with results of an initial
survey at fixed $L_s=8$, permitting a comparison with the $N=2$ data from
\cite{Hands:2016foa}. Since to date U(2$N$) symmetry breaking has not been observed
with DWF, as a warm-up Sec.~\ref{sec:quenched} presents results from a study of
the quenched Thirring model, to see what broken symmetry might look like. We
will see that extrapolation to the $L_s\to\infty$ limit is crucial at the
stronger couplings examined, and introduce an exponential {\em Ansatz\/} which
empirically works well. There does indeed appear to be a range of couplings
where U($2N$) symmetry is broken. Results from an exploratory 
study of a quenched model with the auxiliary formulated just on the domain walls
are also presented. Finally in Sec.~\ref{sec:N=1} we present a
detailed study of the $N=1$ model on $12^3$ at four representative $g^2$
over a range of masses $m$, allowing $L_s$ to vary between 8 and 40 (and in some
cases 48) to facilitate for the first time a controlled $L_s\to\infty$
extrapolation.  Comparison data taken for $N=2$, and for $N=1$ on $16^3$  are
also presented. In addition to the bilinear condensate
$\langle\bar\psi\psi\rangle$, results for correlators of spin-0 mesons will be
given, including the channel with quantum numbers of the would-be Goldstone
bosons. We will argue that at the strongest coupling examined, close to
the effective strong coupling limit, the most plausible explanation of the data
is that U(2) symmetry is spontaneously broken. By contrast, the $L_s$-extrapolated
data for the $N=2$ model in the effective strong coupling limit is consistent with
unbroken symmetry, implying $1<N_c<2$. We summarise and outline plans for future
work in Sec.~\ref{sec:discussion}.

\section{Formulation and Implementation}
\label{sec:formulation}
The most straightforward way to simulate the Thirring model with orthodox
techniques is via introduction of a vector auxiliary field $A_\mu(x)$. The
continuum Lagrangian density is then written
\begin{equation}
{\cal
L}=\bar\psi_i(\partial{\!\!\!\!\,/}+iA{\!\!\!\!
\,/}+m)\psi_i+{N\over{2g^2}}A_\mu^2.
\label{eq:Laux}
\end{equation}
In this form the similarity to an abelian gauge theory is manifest, and it is
clear the Thirring model inherits the same U(2$N$) global symmetry
(\ref{eq:mass},\ref{eq:massless}). The bosonic
action violates the gauge symmetry, however, though this can be remedied by
introduction of a St\"uckelberg scalar leading to a ``hidden local
symmetry''~\cite{Itoh:1994cr}. 
Eqn.~(\ref{eq:Laux}) includes a mass term $m\bar\psi\psi$, which for reducible spinor
representations in 2+1$d$ can be shown to be invariant under parity
inversion. However, it is not unique; as outlined in \cite{Hands:2015qha} there are three possible
parity-invariant mass terms:
\begin{equation}
m_h\bar\psi\psi;
\;\;\;im_3\bar\psi\gamma_3\psi;\;\;\;
im_5\bar\psi\gamma_5\psi;
\label{eq:masses}
\end{equation}
in Euclidean metric the first is hermitian while the two ``twisted'' forms are
antihermitian. Due to their equivalence under U(2$N$) rotations, and the
absence of chiral anomalies in 2+1$d$, they are physically indistinguishable.

In this work fermions are studied on a 2+1+1$d$ lattice using the 
DWF formulation. The action is written~\cite{Hands:2016foa} 
\begin{equation}
S= S_{\rm kin}+S_{\rm int}+S_{\rm bos}
=\sum_{i=1}^N\sum_{x,y}\sum_{s,s^\prime}\bar\Psi_i(x,s){\cal M}(x,s\vert
y,s^\prime)\Psi_i(y,s^\prime)+S_{\rm bos},
\label{eq:action}
\end{equation}
where $\Psi(x,s),\bar\Psi(x,s)$ are defined on the 2+1+1$d$ lattice.
$S_{\rm bos}$ is the action for the auxiliary boson fields $A_\mu(x)$ defined on
links $\mu=0,1,2$, and is an obvious generalisation of the gaussian term in
(\ref{eq:Laux}). For convenience throughout we will use lattice units with
$a\equiv1$, but note here that the dimensionless combination is $g^{-2}a$.
The coordinates $x,y$ denote sites in 2+1$d$, and 
$s$ running along the third direction $x_3$ takes values $1,\ldots,L_s$.
The fermion kinetic action is defined
\begin{equation}
S_{\rm kin}=
\sum_{x,y}\sum_{s,s^\prime}\bar\Psi(x,s)[\delta_{s,s^\prime}D_{Wx,y}
+\delta_{x,y}D_{3s,s^\prime}]
\Psi(y,s^\prime)+m_aS_a.
\label{eq:SDWF}
\end{equation}
$D_W(M)_{x,y}$ is the 2+1$d$ Wilson operator with $M$ the
domain wall height:
\begin{equation}
D_W(M)_{x,y}=-{1\over2}\sum_{\mu=0,1,2}
\left[(1-\gamma_\mu)\delta_{x+\hat\mu,y}+(1+\gamma_\mu)\delta_{x-\hat\mu,y}
\right]
+(3-M)\delta_{x,y},
\label{eq:Ddw}
\end{equation}
while $D_3$ governs hopping along $x_3$:
\begin{equation}
D_{3\,s,s^\prime}
=-\left[P_-\delta_{s+1,s^\prime}
(1-\delta_{s^\prime,L_s})
+P_+\delta_{s-1,s^\prime}(1-\delta_{s^\prime,1})\right]
+\delta_{s,s^\prime}.
\label{eq:D3dw}
\end{equation}
The factors $(1-\delta_{s^\prime,1/L_s})$ implement open boundary conditions at
domain walls located at $s=1,L_s$, while
the projectors $P_\pm\equiv{1\over2}(1\pm\gamma_3)$ also appear
in the identification of target fermions $\psi(x),\bar\psi(x)$ defined
as 2+1$d$ fields localised on the domain walls at $s=1,L_s$:
\begin{equation}
\psi(x)=P_-\Psi(x,1)+P_+\Psi(x,L_s);\;\;\;
\bar\psi(x)=\bar\Psi(x,L_s)P_-+\bar\Psi(x,1)P_+.
\label{eq:4to3}
\end{equation}
The relations (\ref{eq:4to3}) are a major ingredient in the physical
interpretation of DWF, but for now we stress they should be regarded as
assumptions. They permit a definition of the mass term $m_aS_a$ in
(\ref{eq:SDWF}), using (\ref{eq:masses}) with $a=h,3,5$. It is easily checked
that $m_h,m_3$ couple fields defined on opposite walls, while $m_5$ couples fields on
the same wall. 

In Ref.~\cite{Hands:2015dyp} it was shown that in the limit
$L_s\to\infty$ the DWF operator ${\cal M}$ is equivalent to §an 
overlap operator constructed to satisfy 2+1$d$ generalisations of the
Ginsparg-Wilson relations, and \cite{Hands:2015qha}
demonstrated recovery of U($2N$) symmetry in a weakly-coupled
theory, quenched non-compact QED$_3$, in the same limit. Specifically, for
finite $L_s$ the three condensates related by U($2N$) satisfy:
\begin{eqnarray}
{1\over2}\langle\bar\psi\psi\rangle_{L_s}&=&{i\over2}\langle\bar\psi\gamma_3\psi\rangle_{L_S\to\infty}
+\delta_h(L_s)+\epsilon_h(L_s);\\
{i\over2}\langle\bar\psi\gamma_3\psi\rangle_{L_s}&=&{i\over2}\langle\bar\psi\gamma_3\psi\rangle_{L_S\to\infty}
+\epsilon_3(L_s);\label{eq:epsilon3}\\
{i\over2}\langle\bar\psi\gamma_5\psi\rangle_{L_s}&=&{i\over2}\langle\bar\psi\gamma_3\psi\rangle_{L_S\to\infty}
+\epsilon_5(L_s),
\end{eqnarray}
all residuals decaying exponentially with $L_s$
with hierarchy $\delta_h\gg\epsilon_h\gg\epsilon_3,\epsilon_5$. The dominant
residual $\delta_h$ is defined by the imaginary component of the
3-condensate:
\begin{equation}
\delta_h(L_s)=\Im\langle\bar\Psi(1)i\gamma_3\Psi(L_s)\rangle=-\Im\langle\bar\Psi(L_s)i\gamma_3\Psi(1)\rangle,
\label{eq:deltah}
\end{equation}
and is thus measurable even when, as in this work, the action $m_3S_3$ is
used.

\begin{figure}[H]
    \centering
    \includegraphics[width=12.5cm]{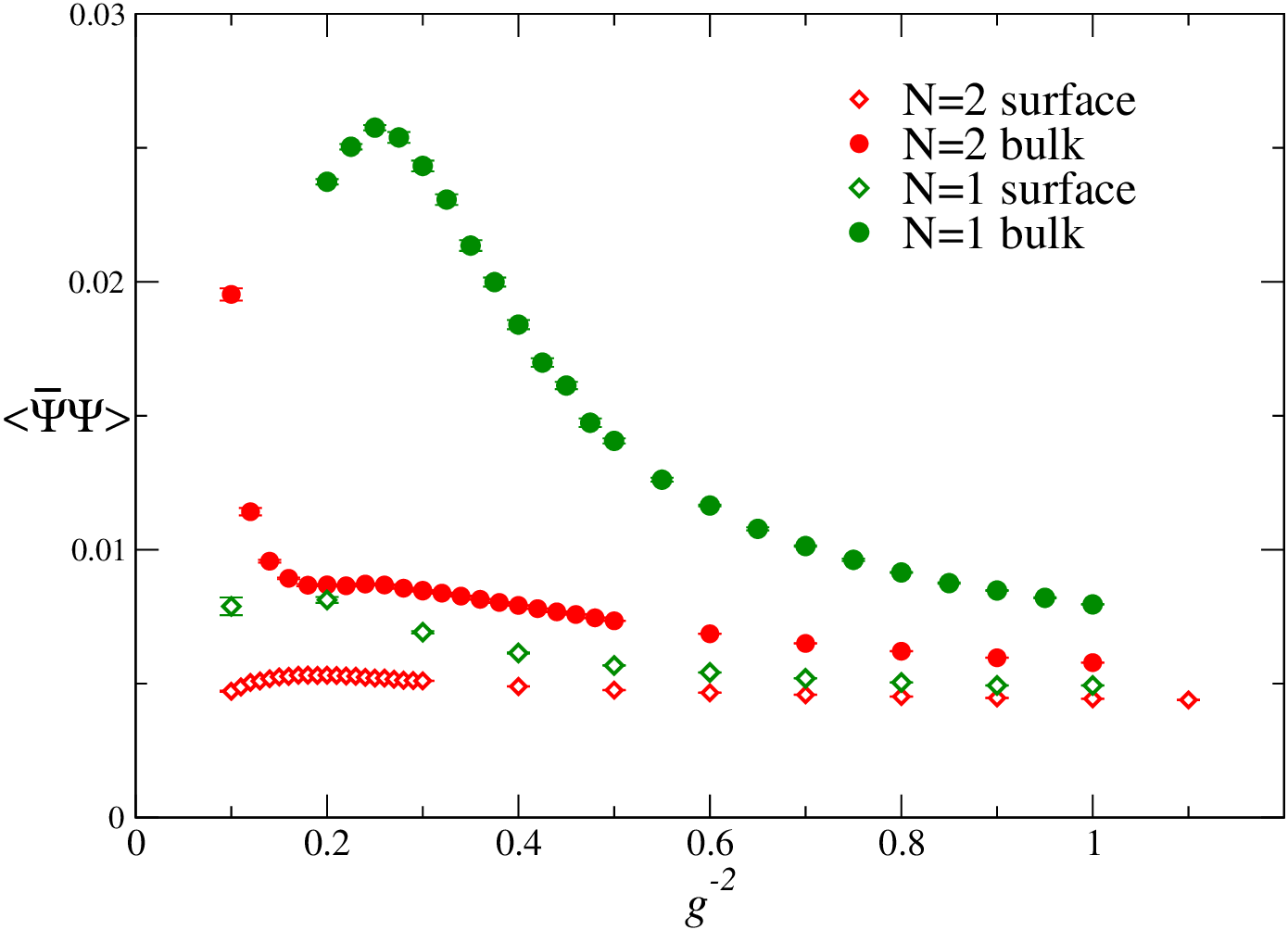}
\caption{(color online) Bilinear condensate $\langle\bar\psi\psi\rangle$ vs. $g^{-2}$ for $N=1,2$ on a
$12^3\times L_s$ lattice with $m=0.01$.}
\label{fig:cond_N}
\end{figure}

To complete the specification of the Thirring model with DWF we need the
fermion-auxiliary interaction $S_{\rm int}$. Two variants were introduced in
Ref.~\cite{Hands:2016foa}. The {\em surface\/} formulation has the link fields
$A_\mu$ linearly §coupled to point-split fermion bilinears defined on the walls:
\begin{eqnarray}
S_{\rm surf}={i\over2}\sum_{x,\mu}A_\mu(x)[\bar\Psi(x,1)\gamma_\mu
P_-\Psi(x+\hat\mu,1)
\!\!\!&+&\!\!\!\bar\Psi(x,L_s)\gamma_\mu P_+\Psi(x+\hat\mu,L_s)]
\label{eq:Thirint}\\
+A_\mu(x-\hat\mu)[\bar\Psi(x,1)\gamma_\mu
P_-\Psi(x-\hat\mu,1)
\!\!\!&+&\!\!\!\bar\Psi(x,L_s)\gamma_\mu P_+\Psi(x-\hat\mu,L_s)].\nonumber
\end{eqnarray}
A similar approach has been adopted in DWF studies of another 2+1$d$ 
theory of interacting fermions, the Gross-Neveu
model~\cite{Vranas:1999nx,Hands:2016foa}. Because the interaction is 
defined only at the walls, (\ref{eq:Thirint}) brings the technical advantage that the Pauli-Villars
determinant needed to formally recover the correct fermion measure as $L_s\to\infty$ does
not depend on $A_\mu$, and hence need not be simulated, making calculations
with (\ref{eq:Thirint}) relatively inexpensive. The {\em bulk\/} formulation
emphasises the resemblence  of the vector auxiliary to a gauge field, defining
a linear interaction between the vector bilinear current and an $s$-independent
$A_\mu$ throughout the bulk:
\begin{equation}
S_{\rm bulk}={i\over2}\sum_{x,\mu,s}A_\mu(x)[\bar\Psi(x,s)(-1+\gamma_\mu)
\Psi(x+\hat\mu,s)]
+A_\mu(x-\hat\mu)[\bar\Psi(x,s)(1+\gamma_\mu)
\Psi(x-\hat\mu,s)].
\label{eq:Thirbulkint}
\end{equation}
Operationally, the fermion operator with $S_{\rm int}=S_{\rm bulk}$ resembles
that of a gauge theory with connection $U_\mu=(1+iA_\mu)$; in other words,
the link field is no longer constrained to be unitary. This choice is not unique
-- other lattice approaches to the Thirring model use unitary link
fields~\cite{Kim:1996xza,Alexandru:2018ddf}
-- but for $N>1$ it ensures there are no fermion interactions higher than
four-point once the auxiliary is integrated out~\cite{DelDebbio:1997dv}.
However, in the same work it was shown that this regularisation fails 
to preserve the transversity of the vacuum polarisation correction to the
$A$-propagator, leading to an additive renormalisation of $g^{-2}$ in the large-$N$
expansion. The consequent uncertainty in identifying the strong-coupling limit
has been explored using staggered fermions in \cite{Christofi:2007ye}, where the
pragmatic approach of identifying the physical strong coupling limit
$g_R^2\to\infty$ with the point where 
$\langle\bar\psi\psi(g^2)\rangle$ has a maximum was found to yield a plausible
equation of state. This point will be further discussed below
Fig.~\ref{fig:cond_N}.

Most of the results presented in the paper are obtained using the bulk
formulation, and it will be shown in Sec.~\ref{sec:mesons} why this is the preferred option. 
Appendix~\ref{app:A} derives some relations for the fermion determinant in the
bulk approach, motivating the use of the RHMC algorithm~\cite{Clark:2003na} for simulation with
$N=1$. For even $N$, the HMC algorithm outlined in \cite{Hands:2016foa} is
sufficient.
The pseudofermion action used in RHMC is 
\begin{equation}
S=\sum_{i=1}^N\sum_{x,y,s,s^\prime}
\Phi_i^\dagger(x,s)\left\{[{\cal M^\dagger M}_{m_h=1}]^{N\over4}
[{\cal M^\dagger M}_{m_3=m}]^{-{N\over2}}[{\cal M^\dagger
M}_{m_h=1}]^{N\over4}\right\}_{x,s\vert y,s^\prime}\Phi_i(y,s^\prime),
\label{eq:pf}
\end{equation}
where subscripts on ${\cal M^\dagger M}$ denote non-vanishing mass terms.
We choose the domain wall mass $M=1$.
The components with $m_h=1$ describe Pauli-Villars fields needed to cancel bulk
contributions to the determinant and ensure coincidence with the correct overlap
operator as
$L_s\to\infty$~\cite{Hands:2015dyp}. The fractional matrix powers 
needed to compute (\ref{eq:pf}) are estimated by means of a rational
approximation
\begin{equation}
({\cal A})^p\simeq r_p({\cal A})=\alpha_0+\sum_{i=1}^{N_{pf}}{\alpha_i\over{{\cal
A}+\beta_i}}
\label{eq:PF}
\end{equation}
The required coefficents $\alpha,\beta$ are calculated with the Remez algorithm
using the implementation available at \cite{ClarkKennedy}. They were chosen such
that over a spectral range $(0.0001,50)$ (which accommodates the upper limit 
$(2(2+1+1)-M)^2$ obtained in the free-field limit of (\ref{eq:action})), 
$\vert r_p(x)-x^p\vert$  is less than
$10^{-6}$ for matrices needed during guidance and $10^{-13}$ for those needed in
the Hamiltonian calculations required for the acceptance step of the algorithm. 
This appears to be a conservative requirement for the systems studied to date,
and translates into $N_{pf}=12$ (guidance) and $N_{pf}=25$ (acceptance).
The need for further refinement cannot be ruled out for future studies of critical systems.

The partial fraction expansion (\ref{eq:PF}) is efficiently calculated using the
multi-shift procedure described in \cite{Frommer:1995ik}, which in turn requires
the use of a hermitian Lanczos solver such as described in ~\cite{Golub}. For the
systems we have examined, particularly as $L_s$ is made large, maintaining
orthonormality of the Lanczos vectors generated at each successive iteration
requires double precision arithmetic; on the same systems the conjugate
gradient algorithm used in measurement routines runs happily in single
precision. For an  evaluation of $x=({\cal M^\dagger M})^p\Phi$, the convergence
criterion adopted is 
\begin{equation}
\max_i\vert\alpha_i\rho_i\vert<{{4L_sV\varepsilon^2}\over{N_{pf}\vert\Phi\vert}}
\end{equation}
where $\rho_i$~\cite{Golub} is a real variable parametrising the magnitude of the latest
increment to the solution vector $x_i$ (where  
$x=\alpha_0\Phi+\alpha_ix_i)$, and $\varepsilon=10^{-6}$ (guidance) and $10^{-9}$
(acceptance).
Finally, it should be noted that the matrix inversions are numerically demanding,
especially as the coupling becomes strong, possibly as a consequence of the 
non-unitarity of the link felds $U_\mu$. For instance, on the largest 
$16^3\times40$ volume studied, at the strongest coupling $g^{-2}=0.3$ and the smallest
mass $m=0.01$, the Lanczos
solver in the Hamiltonian calculation requires roughly 11000 iterations to
achieve convergence. The
conjugate gradient solver operating on stochastic noise sources in the measurement routine 
on the same system 
requires roughly 4700 iterations.
 
\begin{figure}[H]
    \centering
    \includegraphics[width=12.5cm]{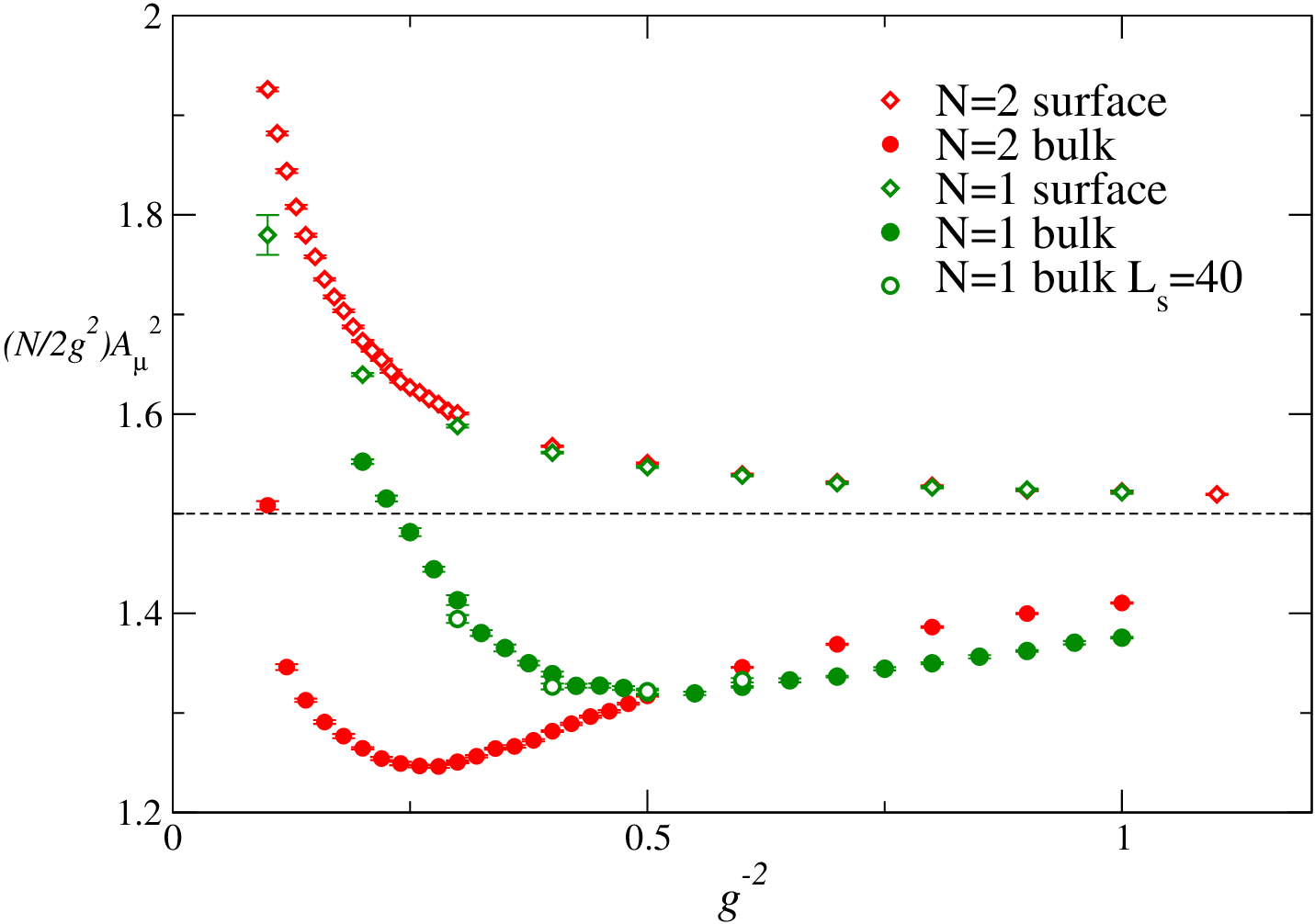}
\caption{(color online) Bose action density ${N\over{2g^2}}\langle A_\mu^2\rangle$ vs. $g^{-2}$ for
$N=1,2$.} 
\label{fig:bose_N}
\end{figure}
For orientation, in the remainder of this section we present the main features of the model based
on simulations with fixed finite $L_s$. All results are taken using the $m_3S_3$ mass term, and
henceforth for convenience when
there is no possibility of confusion
we will often write the associated condensate 
$\langle\bar\psi i\gamma_3\psi\rangle$ as $\langle\bar\psi\psi\rangle$. 
Fig.~\ref{fig:cond_N} shows $\langle\bar\psi\psi\rangle$ vs $g^{-2}$  for
$m=0.01$ on $12^3\times L_s$, for both surface and bulk formulations with
$N=1$ ($L_s=8$) and $N=2$ ($L_s=16$). The
$N=2$ results obtained using the HMC algorithm were
first presented in \cite{Hands:2016foa}.

In all cases the condensate increases as the coupling is increased from weak to
strong, until it reaches a maximum in the region $g^{-2}\approx0.2$ -- 0.3.
This non-monotonic behaviour maximum is also observed in simulations using both
staggered
~\cite{DelDebbio:1997dv,Christofi:2007ye} and SLAC~\cite{Schmidt:2016rtz} fermions, and is
associated with strong-coupling artifacts possibly due to the non-transversity of the
auxiliary propagator discussed above; following \cite{Christofi:2007ye} we will identify the maximum with the
approximate location of the effective strong coupling limit, and focus our attention on the
weak-coupling side of this maximum.  
With the vertical scale chosen to accommodate the $N=1$ bulk data, the
condensates obtained with the surface formulation in this region are very small and show
little dependence on coupling. The bulk formulation yields larger condensates,
but the most striking feature of Fig.~\ref{fig:cond_N} is the sharp rise in the
bulk condensate for
$g^{-2}<0.6$ for $N=1$; it is already apparent that the tendency for fermions
and antifermions to pair is more significant here than for any
case previously examined. 

Fig.~\ref{fig:bose_N} shows the auxiliary action density
${N\over{2V}}g^{-2}\sum_{x\mu}
A_\mu^2(x)$ vs. $g^{-2}$ on the same systems. 
This measurable offers
an interesting diagnostic of the UV
properties of the different model approaches. 
First note that in the continuum, a comparison of $\partial\ln
Z/\partial g^2$ obtained using the original 
action (\ref{eq:L}) and with the bosonised form (\ref{eq:Laux}), following a
change in functional integration variables, results in the
following identity for the boson action:
\begin{equation}
{N\over{2g^2}}\langle
A_\mu^2\rangle={3\over2}+{g^2\over{2N}}\langle(\bar\psi i\gamma_\mu\psi)^2\rangle.
\label{eq:bosac}
\end{equation}
Assuming smooth behaviour of the expectation of the square of the fermion
current on the RHS of (\ref{eq:bosac}),
we therefore expect departures from the free-field value $3\over2$ to increase monotonically
with $g^2$, and Fig.~\ref{fig:bose_N} shows this is indeed the case
for the surface model with $N=1,2$. For the bulk model the corresponding
relation contains terms of the form
$\bar\Psi\gamma_\mu\Psi(s)\bar\Psi\gamma_\mu\Psi(s^\prime)$,  
$\bar\Psi\gamma_\mu\Psi(s)\Phi^\dagger\gamma_\mu\Phi(s^\prime)$,  
with $s,s^\prime=1,\ldots,L_s$ and $\Phi,\Phi^\dagger$ are Pauli-Villars fields, 
ie. there are contributions from 
bulk fields whose interpretation is not as transparent. In fact, Fig.~\ref{fig:bose_N}
shows the correction to the free-field result has the opposite sign except at
the very strongest coupling.

The contrast between surface and bulk formulations was already
noted in \cite{Hands:2016foa}. Here we note that $N=1,2$ yield very similar
results for the surface formulation, but for $N=1$ there is a marked contrast
in the bulk results once $g^{-2}\lapprox0.5$, again hinting at interesting
strong coupling behaviour. Fig.~\ref{fig:bose_N} also plots $N=1$ bulk data from
$12^3\times40$ showing small but significant disparities; this is a reminder of the
importance of seeking the $L_s\to\infty$ limit of all observables in the DWF
approach, in particular the bilinear condensate. Sec.~\ref{sec:quenched} presents 
a first investigation in this direction in the quenched limit $N=0$, and enables us to
address the question {\em what does spontaneous symmetry breaking due to
bilinear condensation look like with DWF?\/}. 

\section{Results in the Quenched Limit $N=0$}
\label{sec:quenched}

\begin{figure}[H]
    \centering
    \includegraphics[width=12.5cm]{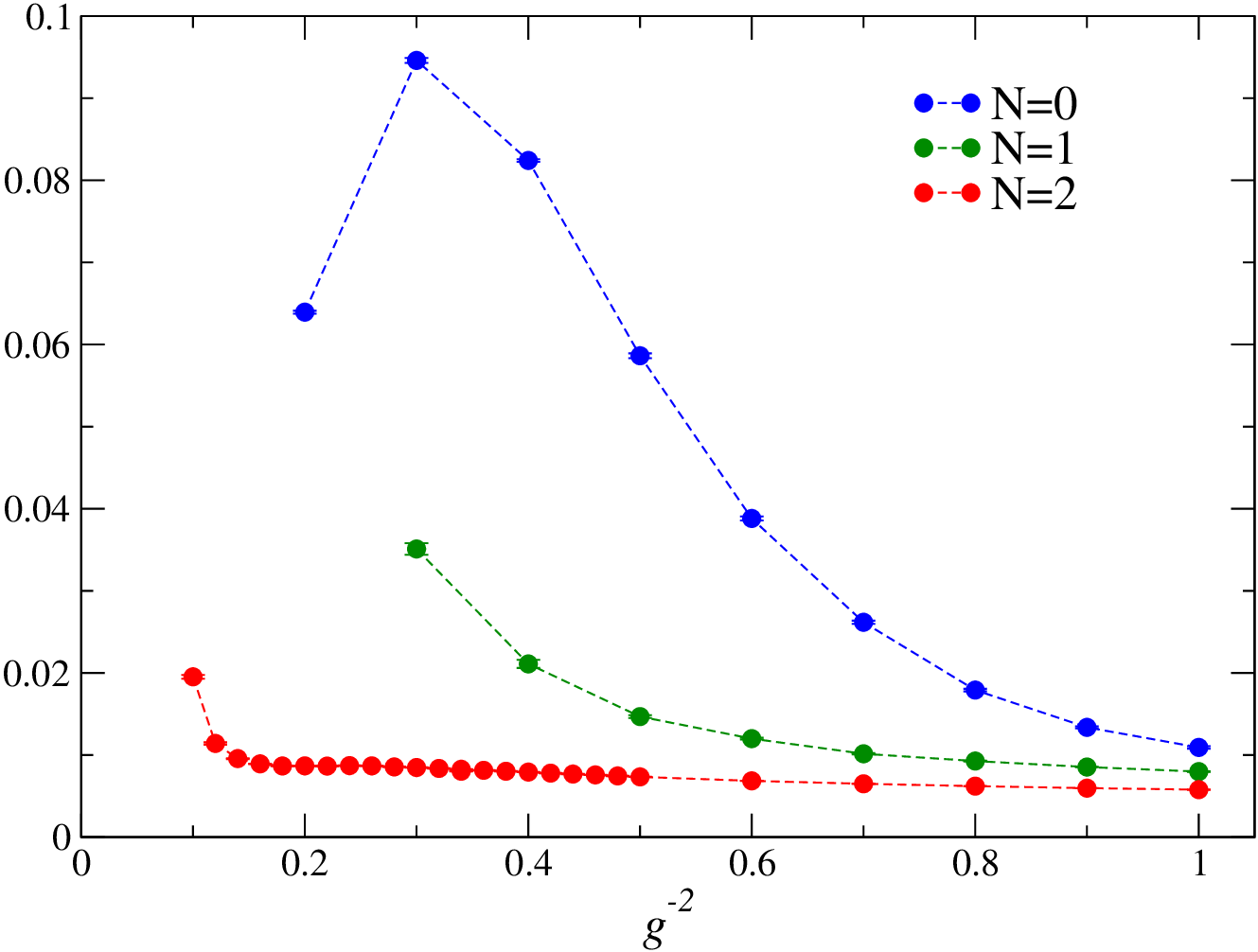}
\caption{(color online) $\langle\bar\psi\psi\rangle$ vs. $g^{-2}$ for
$N=0,1,2$ on $L^3\times16$ with $m=0.01$.}
\label{fig:N_012}
\end{figure}

The quenched theory with $N=0$ is technically very simple to explore; one simply
performs fermionic measurements using the operator ${\cal M}$ on field
configurations inexpensively generated using the gaussian auxiliary action.
Unlike gauge theory, there is no theoretical expectation that the results have 
any relevance to the full theory; this is best understood via 
the auxiliary propagator $S_A(x)$ in the large-$N$ expansion, 
which at strong coupling decays as $\vert x\vert^{-2}$ 
as a result of vacuum polarisation corrections~\cite{Hands:1994kb}, but which remains a
contact $S_A(x)\sim\delta^d(x)$ in the quenched limit. Fig.~\ref{fig:N_012}
compares condensate data obtained with the bulk formulation 
for $N=0,1,2$ for $m=0.01$ with $L_s=16$; for $N>0$ the 
spacetime volume is $12^3$, but the low computational cost enabled the
quenched study on $16^3$. Compared to Fig.~\ref{fig:cond_N} the vertical scale has been extended
to accommodate the quenched data: the hierarchy
$\langle\bar\psi\psi(N=0)\rangle\gg\langle\bar\psi\psi(N=1)\rangle\gg\langle\bar\psi\psi(N=2)\rangle$
is as expected; the low eigenvalues of the effective Dirac operator
responsible for the condensate signal via the Banks-Casher relation also
suppress the determinant in the path integral measure.

To explore the $L_s\to\infty$ limit we performed a systematic study of
the bilinear condensate $\langle\bar\psi\psi(g^2,m)\rangle$ 
on $16^3\times L_s$ 
with $L_s=8,\ldots,40$, $g^{-2}=0.2,0.3,\ldots,1.0$ and
$m=0.01,0.02,\ldots,0.05$. Each boson configuration, separated by 100 HMC
trajectories, was analysed using 10 stochastic noise vectors located on either
wall. 25000 trajectories were studied for $L_s=8,16$, and 5000 for
$L_s=24,32,40$. 

\begin{figure}[H]
    \centering
    \includegraphics[width=12.5cm]{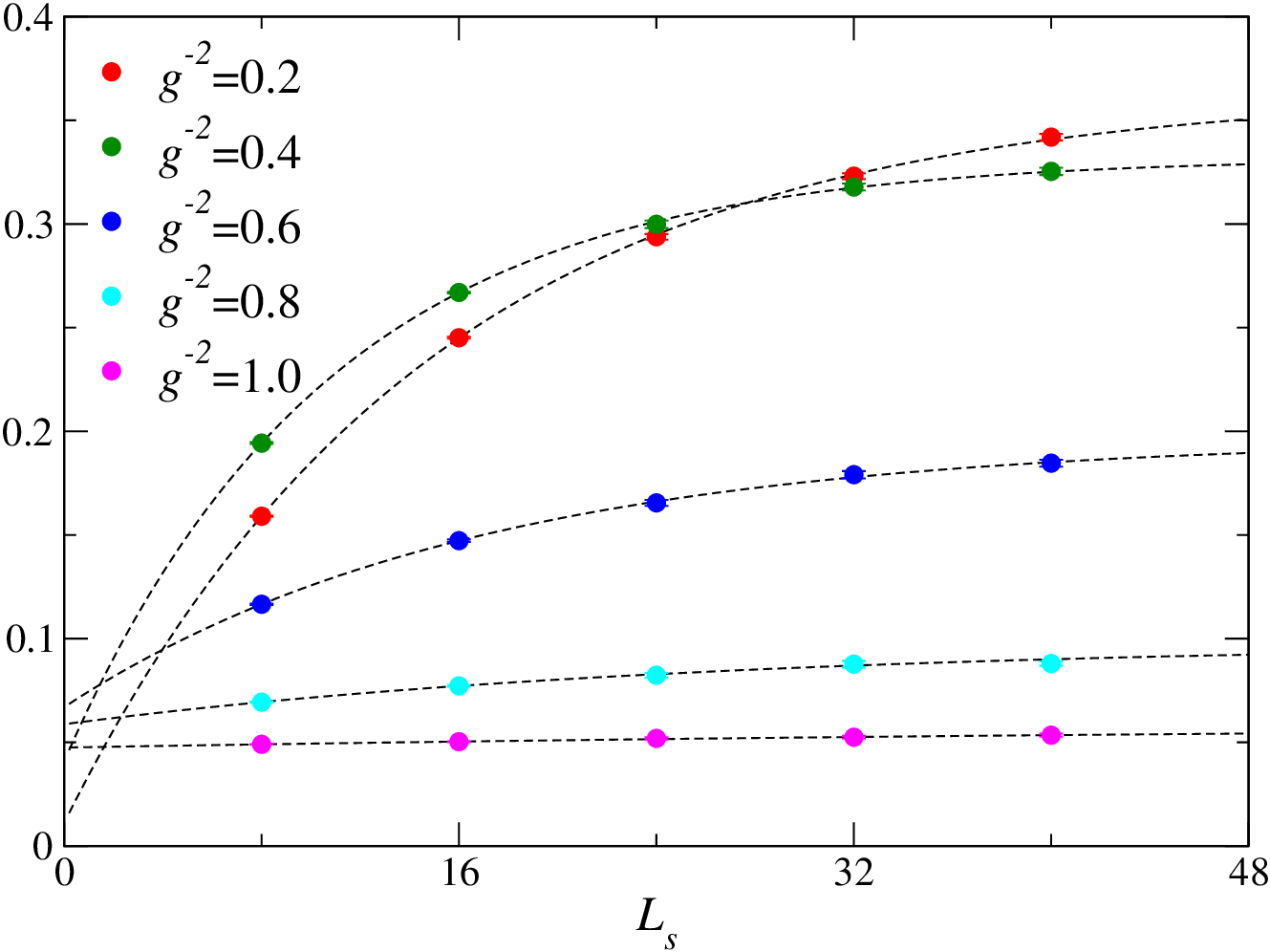}
\caption{(color online) $\langle\bar\psi\psi\rangle$ vs. $L_s$ for various $g^{-2}$ with
$m=0.05$.}
\label{fig:cond_m05}
\end{figure}

\begin{figure}[H]
    \centering
    \includegraphics[width=12.5cm]{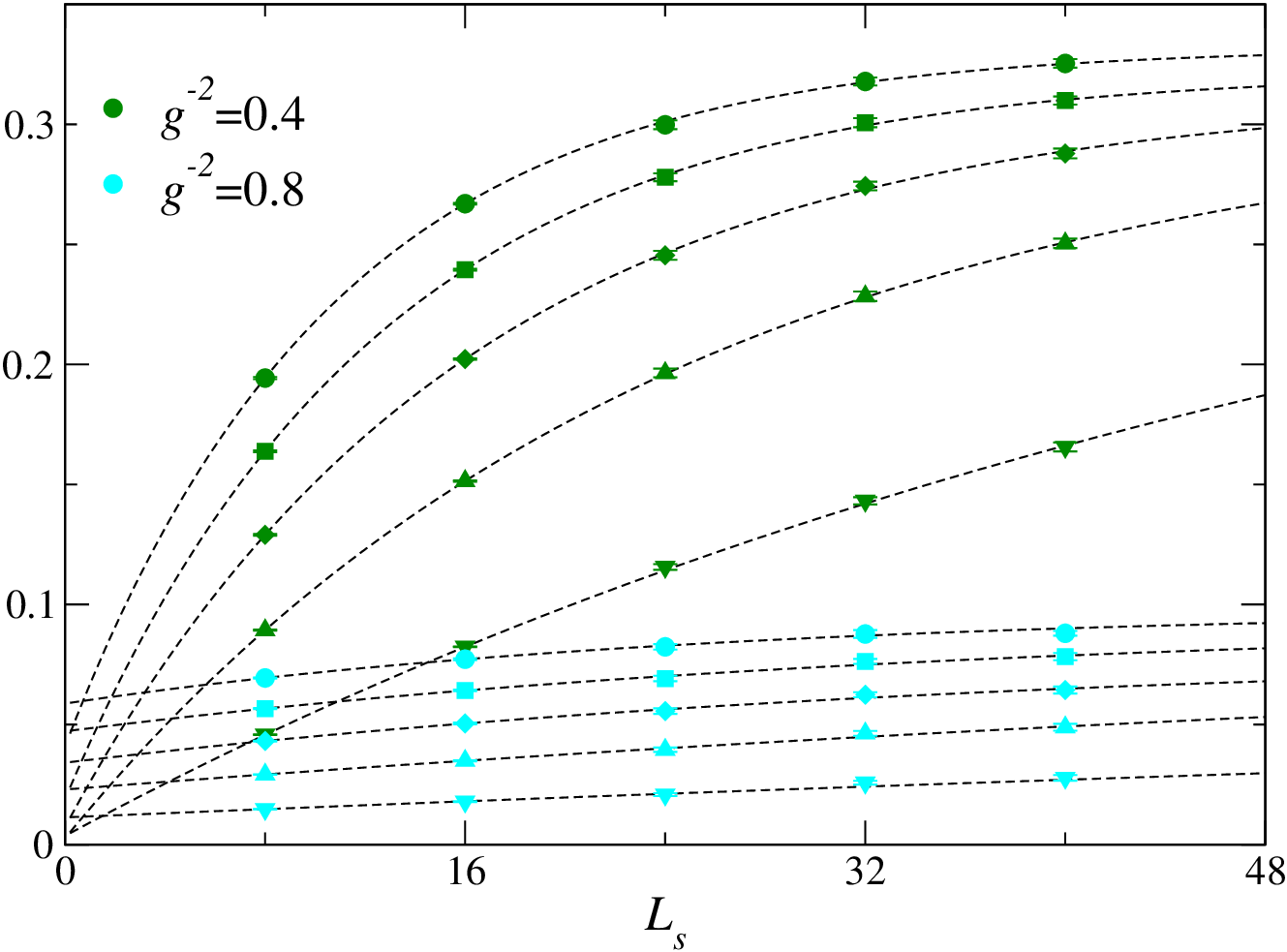}
\caption{(color online) $\langle\bar\psi\psi\rangle$ vs. $L_s$ for $m=0.01,\ldots0.05$ with
$g^{-2}=0.4,0.8$.}
\label{fig:b04vsb08}
\end{figure}

Results for $\langle\bar\psi\psi(L_s)\rangle$ with $m=0.05$ and varying $g^{-2}$ are shown in
Fig.~\ref{fig:cond_m05}, and for varying $m$ at $g^{-2}=0.4, 0.8$ in
Fig.~\ref{fig:b04vsb08}. It is evident that finite-$L_s$ corrections are
significant, and increase in importance as the coupling grows.
We have modelled them using the notation of (\ref{eq:epsilon3})  as follows:
\begin{equation}
\langle\bar\psi\psi\rangle_{L_s=\infty}-\langle\bar\psi\psi\rangle_{L_s}
=2\epsilon_3(L_s,m,g^2)=A(m,g^2)e^{-\Delta(m,g^2)L_s}.
\label{eq:expfit}
\end{equation}
The resulting three parameter fits are plotted as dashed lines in
Figs.~\ref{fig:cond_m05},\ref{fig:b04vsb08}. The exponential form
(\ref{eq:expfit}) works well across the dataset, but the asymptotic value
$\langle\bar\psi\psi\rangle_\infty$ becomes
poorly constrained as $m\to0$ resulting in large uncertainties in this limit.
Also note that the strongest coupling $g^{-2}=0.2$ looks to be an outlier in
both Figs. \ref{fig:N_012} and \ref{fig:cond_m05}, reflecting the probable
influence of strong coupling artifacts.

\begin{figure}[H]
    \centering
    \includegraphics[width=12.5cm]{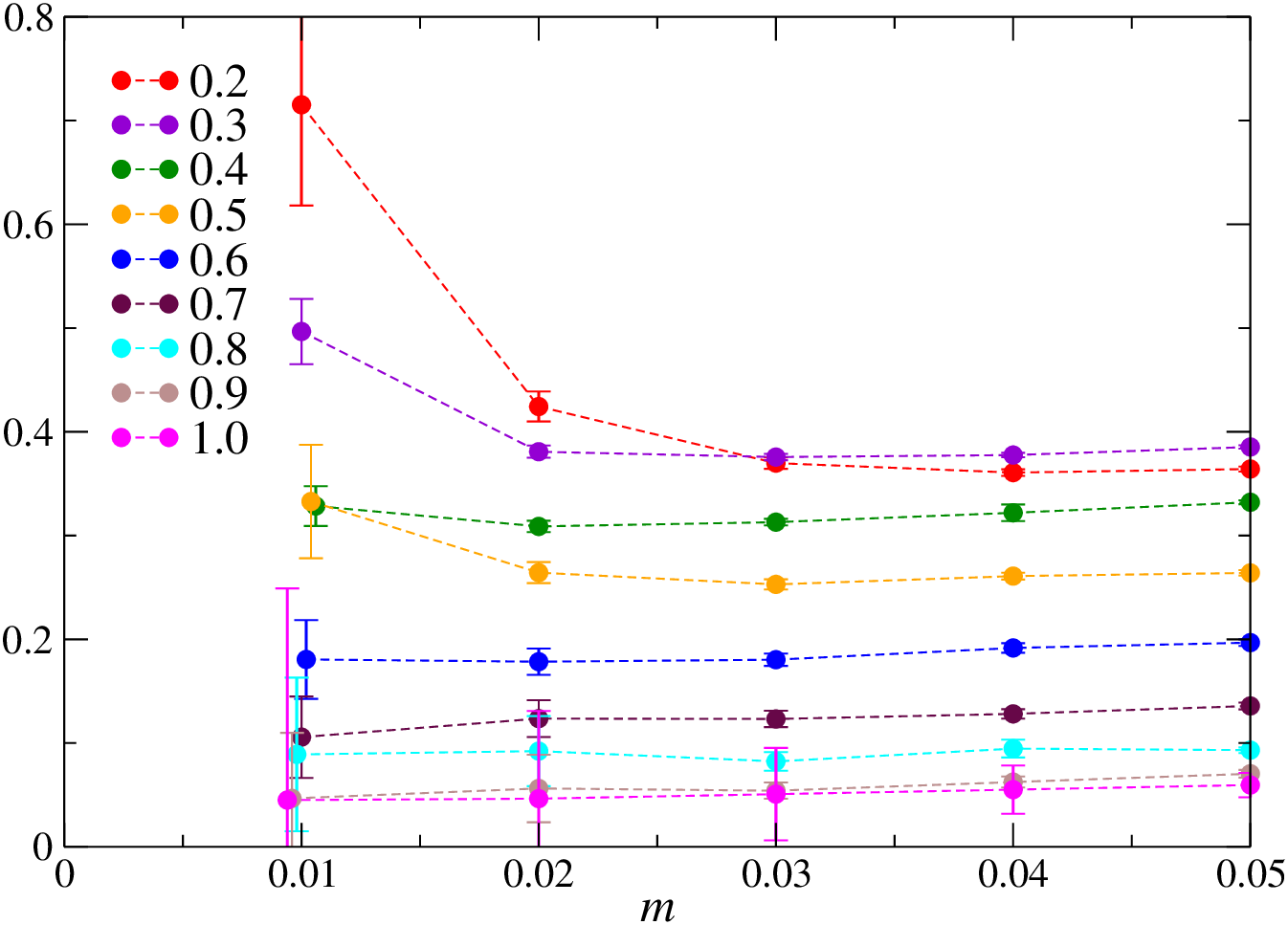}
\caption{(color online) $\langle\bar\psi\psi\rangle_\infty$ vs. $m$ for $g^{-2}\in[0.2,1.0]$.
For clarity some points with large errorbars have been horizontally displaced.}
\label{fig:Lsinfty}
\end{figure}
Results for $\langle\bar\psi\psi\rangle_\infty$ obtained using fits to
(\ref{eq:expfit}) are plotted for various $g^{-2}$ as a function of mass $m$ in Fig.~\ref{fig:Lsinfty}. 
The curves are for the most part remarkably $m$-independent. The
$m\to0$ limit of the strong coupling data at  $g^{-2}=0.2$ and possibly 0.3  are
affected by artifacts as discussed above, and the large errorbars in the same limit for
$g^{-2}\gapprox0.8$ reflect poorly constrained fit parameters associated
with the
lack of curvature seen in Fig.~\ref{fig:b04vsb08}. It is difficult to say
anything definitive in either case. However. Fig.~\ref{fig:Lsinfty} supports a coupling
window $g^{-2}\in(0.4,0.7)$ where $\lim_{m\to0}\langle\bar\psi\psi\rangle_\infty$ is plausibly
non-zero, implying broken U(2$N$) symmetry. We therefore deduce the existence of
U(2$N$) symmetry breaking for $N=0$ for sufficiently strong coupling, though the
data is not of sufficient quality to determine whether there is a critical
$g_c^2$ such that symmetry is restored for $g^2<g_c^2$, such as occurs in quenched
QED$_4$ with staggered fermions~\cite{Kocic:1990fq}. Nonetheless, this exercise
suggests that $N_c>0$ for the Thirring model.

\begin{figure}[H]
    \centering
    \includegraphics[width=12.5cm]{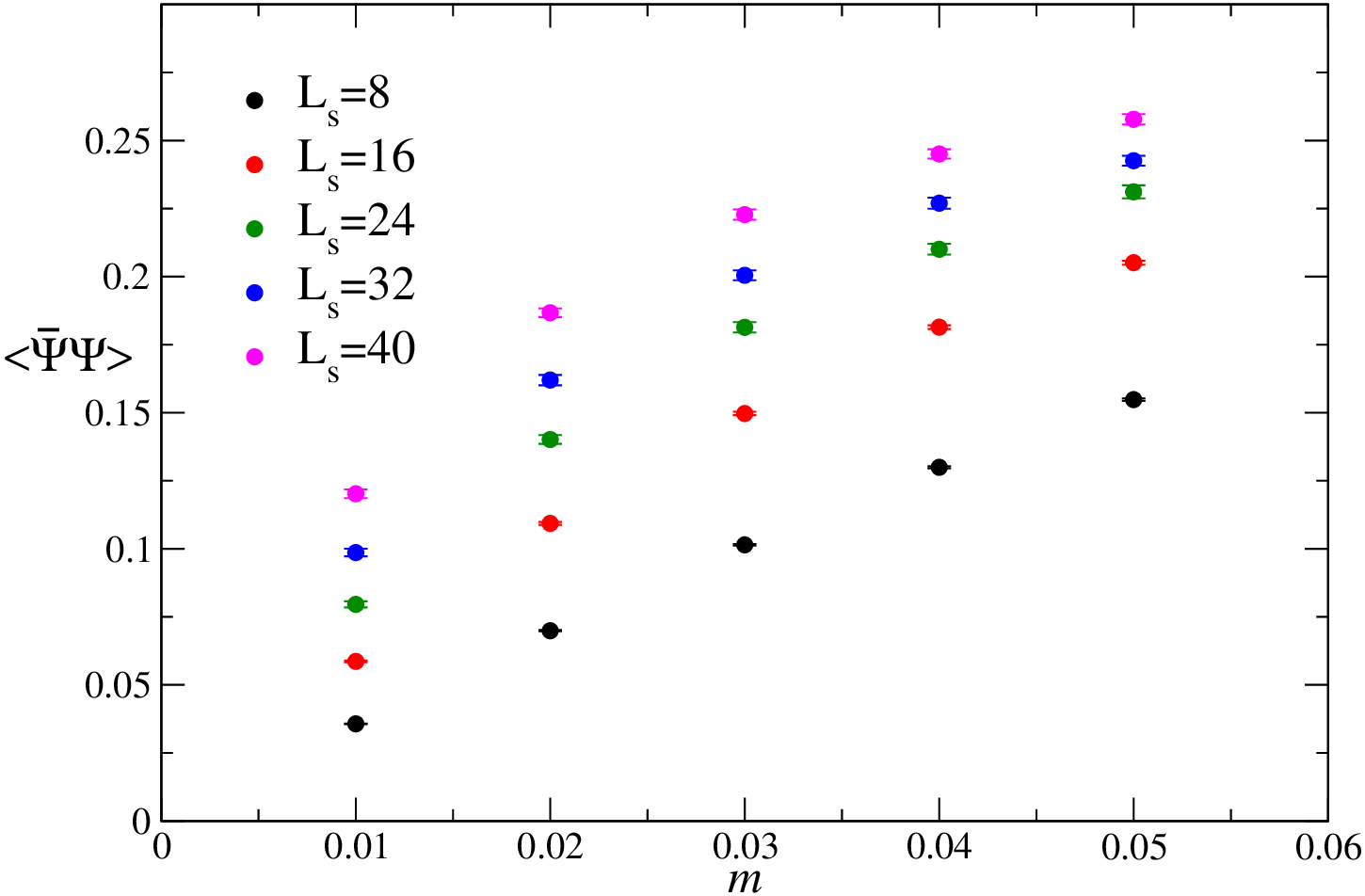}
\caption{(color online) $\langle\bar\psi\psi\rangle_{L_s}$ vs. $m$ for
$g^{-2}=0.4$.}
\label{fig:fixedLs}
\end{figure}

To emphasise the importance of first taking the $L_s\to\infty$ limit,
in Fig.~\ref{fig:fixedLs} we plot the $g^{-2}=0.4$
data vs. $m$ at fixed $L_s$. While the increasing curvature of the data with
$L_s$ is suggestive, there is no compelling evidence to support a
non-zero intercept on the vertical axis as $m\to0$. The $m\to0$ and
$L_s\to\infty$ limits do not commute.
 
\begin{figure}[H]
    \centering
    \includegraphics[width=12.5cm]{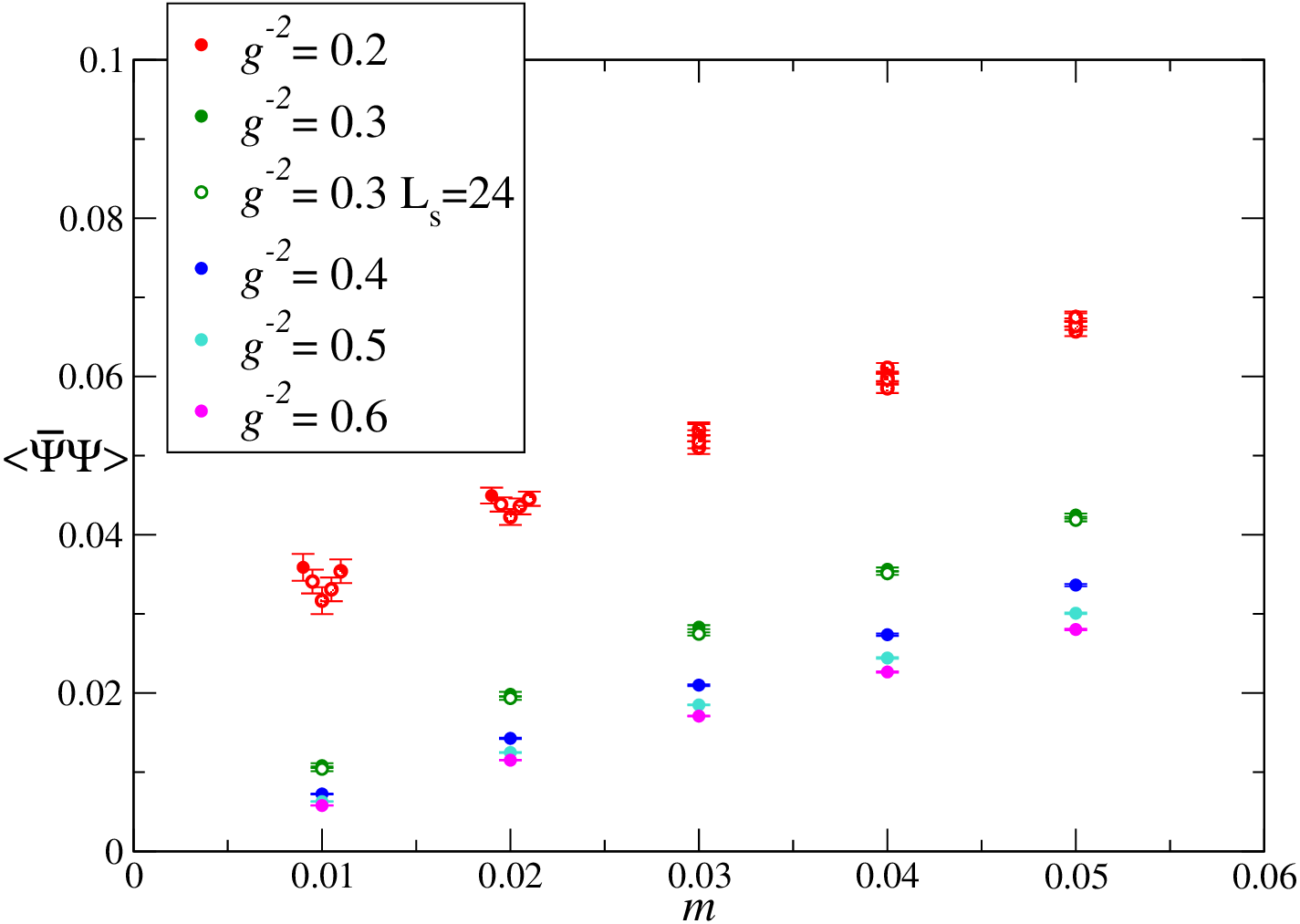}
\caption{(color online) $\langle\bar\psi\psi\rangle$ vs. $m$ for
various $g^{-2}$,$L_s$ for the quenched surface model. The differing symbols
for $g^{-2}=0.2$ denote $L_s=8,16,\dots40$}
\label{fig:N=0_surf}
\end{figure}
For completeness, in Fig.~\ref{fig:N=0_surf} we plot bilinear condensate data
for the quenched surface model. Data have been taken at couplings ranging
from the relatively weak $g^{-2}=0.6$ to $g^{-2}=0.2$ on $16^3\times L_s$, with
$L_s=8$ at all couplings, increasing up to 24 ($g^{-2}=0.3$) and 40
($g^{-2}=0.2$), using  at least 12500 HMC trajectories in all cases. 
The abciss\ae\/ of some datapoints at this strongest coupling have been
slightly displaced for clarity. The emerging picture is qualitatively different from
the bulk case; here there is no evidence for any systematic change in the signal as $L_s$
is increased. For $g^{-2}\geq0.3$ the data show a fairly weak
$g^2$-dependence with $\lim_{m\to0}\langle\bar\psi\psi\rangle=0$ consistent
with the absence of symmetry breaking. By contrast data at $g^{-2}=0.2$ admit a plausible
extrapolation to a non-zero intercept at $m=0$, implying condensation at this
strongest coupling, and suggesting both that $N_c>0$, and that there exists a
critical $g_c^{-2}>0$ at which the symmetry is restored.
Disentangling these effects from the strong-coupling artifacts discussed below
Fig.~\ref{fig:cond_N} would require an extensive programme of further simulations.
To summarize, the quenched exercise shows the importance, at
least for the bulk formulation,  of taking data at
varying $L_s$ and performing a plausible extrapolation to the U($2N$)-symmetric
limit $L_s\to\infty$ using the exponential {\em Ansatz\/} (\ref{eq:expfit}).
While the
quenched limit does not correspond to a unitary field theory (the
$m$-independence of
the curves in Fig.~\ref{fig:Lsinfty} may be a symptom of this), these results
are encouraging because they illustrate at least the possibility of finding 
symmetry breaking in the Thirring model with DWF. We will apply the same
procedure to $N=1$ in the next Section.

\section{Results for $N=1$}
\label{sec:N=1}
Next we present results from simulations of the full field theory with $N=1$,
with emphasis on the $L_s\to\infty$ limit. 
As outlined in Sec.~\ref{sec:formulation}, the required RHMC simulations are
numerically demanding, so the study has been limited to four couplings
$g^{-2}=0.3,0.4,0.5,0.6$  chosen to span
the region of greatest variation in Fig.~\ref{fig:cond_N} while
remaining on the weak-coupling side of the maximum. 
Data were taken at each of 5 masses $m=0.01,\ldots,0.05$, 
and unless stated on a $12^3$ spacetime lattice. 
To probe the large-$L_s$
limit we examined $L_s=8,16,24,32$ and 40, except at the strongest coupling
$g^{-2}=0.3$, where runs with $L_s=48$ were also performed for the three lightest
masses. 
To explore potential volume
effects we also simulated a $16^3$ lattice at $g^{-2}=0.3,0.6$ with $m=0.01$.
For each parameter set a
minimum of 600 RHMC trajectories of mean length 1.0 were generated, with
observables calculated every five trajectories. 

\subsection{Bilinear Condensate $\langle\bar\psi\psi\rangle$}
\label{sec:condensate}

\begin{figure}[H]
    \centering
    \includegraphics[width=12.5cm]{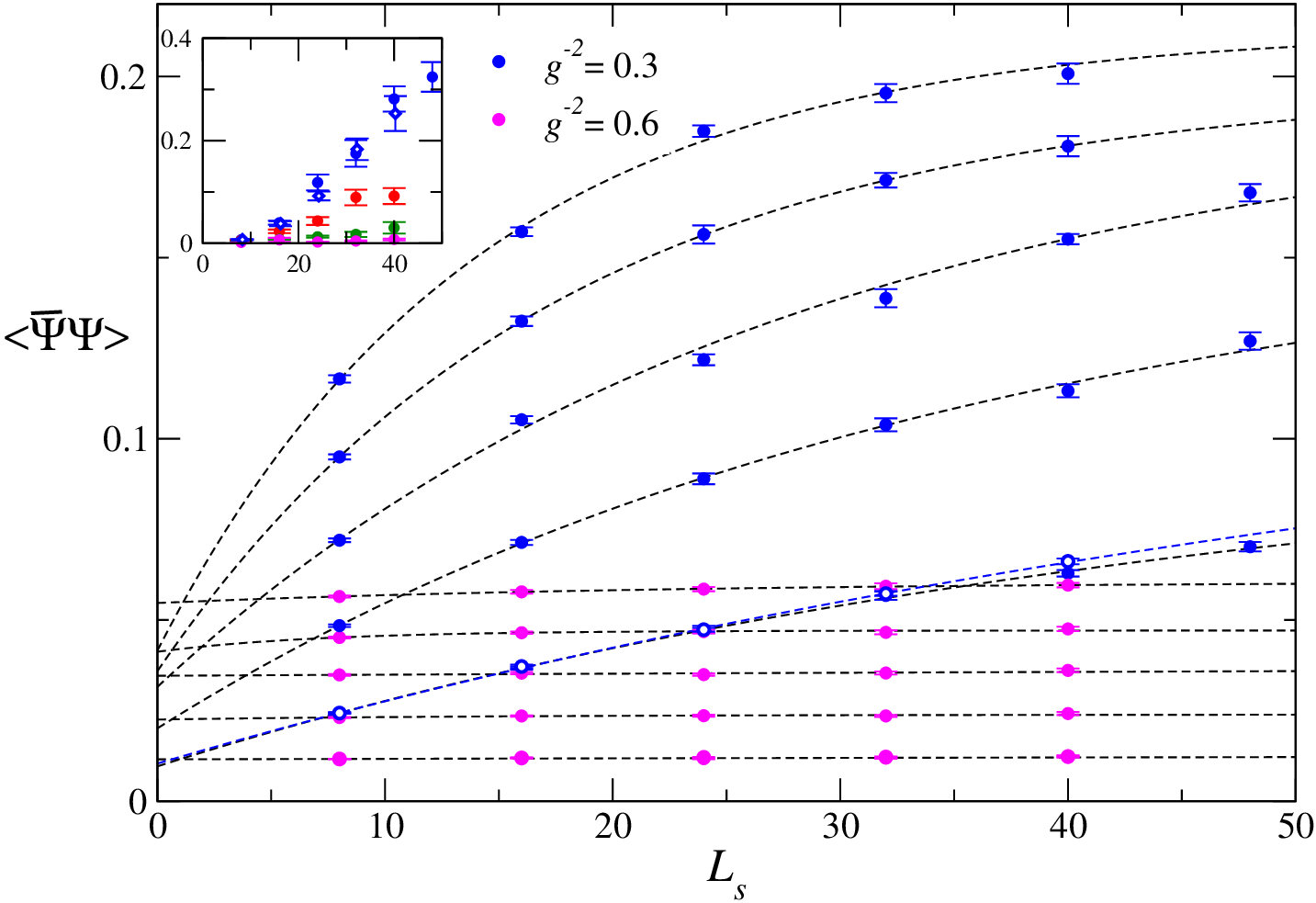}
\caption{(color online) $\langle\bar\psi\psi\rangle$ vs. $L_s$ for
$g^{-2}=0.3,0.6$. Open symbols correspond to $16^3$, and dashed lines fits to
(\ref{eq:expfit}). The inset plots
$\langle(\bar\psi\psi)^2\rangle-\langle\bar\psi\psi\rangle^2$ vs. $L_s$ for
various $g^{-2}$; the colour code is that of Fig.~\ref{fig:N_1}.}
\label{fig:comparison}
\end{figure}

Just as in the quenched case, the importance of finite-$L_s$ corrections
increases markedly as the coupling gets stronger. Fig.~\ref{fig:comparison}
compares data taken at the strongest and weakest couplings explored for all
five mass values, and bears a striking resemblance to Fig.~\ref{fig:b04vsb08}. 
Indeed, the {\em Ansatz} (\ref{eq:expfit}) again
gives a very good description of the condensate data, with $\chi^2$
per degree of freedom for each fit usually $\lapprox2$ across the entire dataset. In what
follows the condensate values extrapolated to $L_s\to\infty$ are based on all
available data, with no points excluded from the fit. Fig.~\ref{fig:comparison} also includes
results taken on $16^3$ denoted by open symbols. On the scale of the plot,
volume effects are only discernible at strong couplings and the largest
available $L_s$; fits from both $12^3$ and $16^3$ will be presented
below.

The inset of Fig.~\ref{fig:comparison} shows the variance of
$\langle\bar\psi\psi\rangle$, or in physical terms the disconnected contribution
to the longitudinal susceptibility, as a function of $L_s$. This demonstrates that the
$L_s\to\infty$ limit is also key to characterising the 
fluctuations of a would-be order parameter;  indeed it is clear that still
larger $L_s$ will
be needed before this observable converges, particularly at stronger couplings.
Note that data from $16^3$ are compatible with $12^3$,
demonstrating that the observed growth is a finite-$L_s$ artifact and is not 
associated with critical fluctuations.

\begin{figure}[H]
    \centering
    \includegraphics[width=12.5cm]{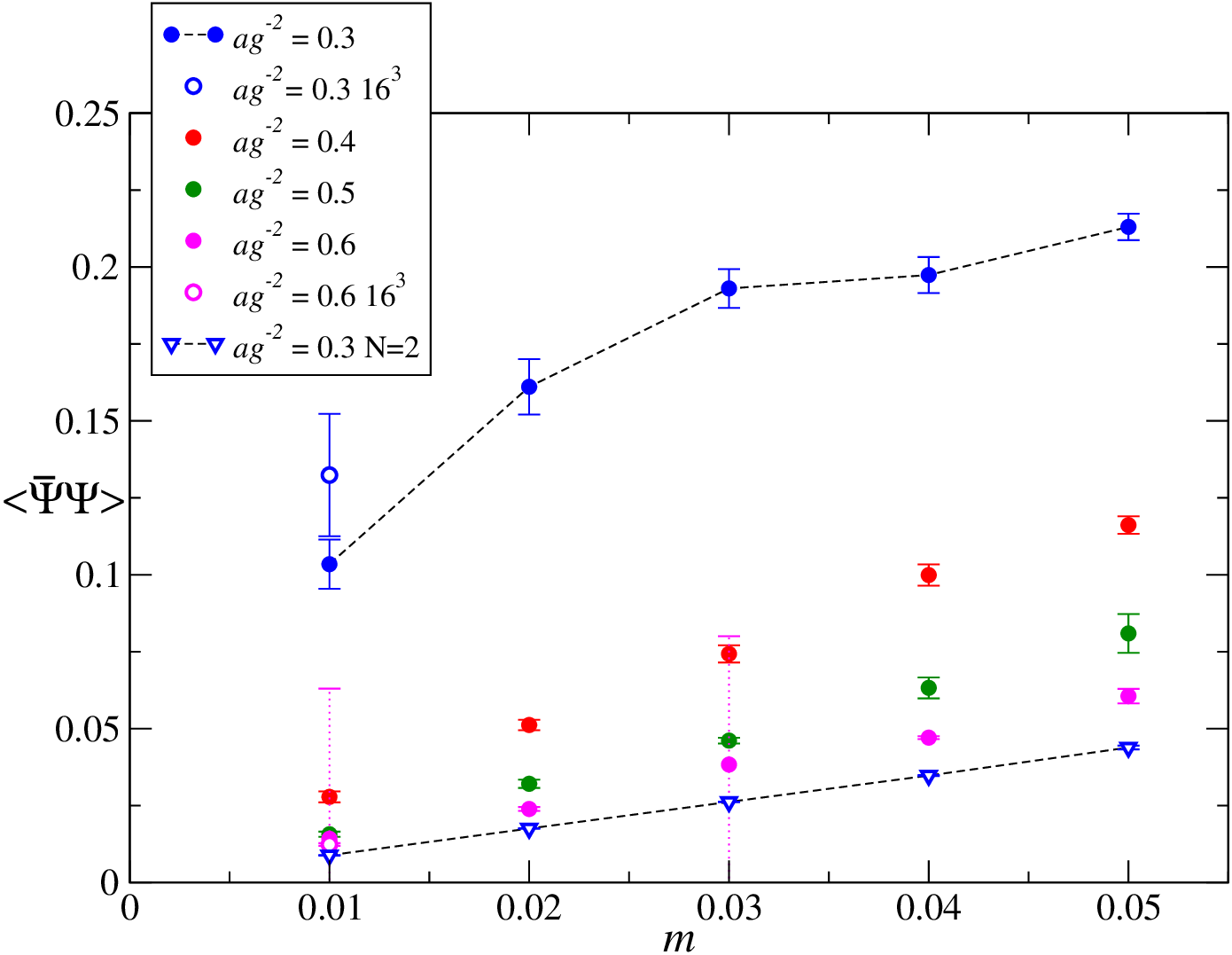}
\caption{(color online) Bilinear condensate $\langle\bar\psi\psi(m)\rangle$ obtained for
various $g^{-2}$ in the
$L_s\to\infty$ limit. 
Open circles denote data from $16^3$.
Open triangles denote data from the Thirring model with $N=2$}
\label{fig:N_1}
\end{figure}

Fig.~\ref{fig:N_1} shows the bilinear condensate
$\langle\bar\psi\psi\rangle$ following the $L_s\to\infty$ limit obtained using
(\ref{eq:expfit}). Different colours correspond to different couplings -- note
that uncertainties in the $L_s\to\infty$ extrapolation occasionally result in
very large errorbars at the weakest coupling $g^{-2}=0.6$.  For
$g^{-2}\geq0.4$, the data are consistent with the behaviour
$\langle\bar\psi\psi(m)\rangle\propto m$, implying no symmetry breaking in the
limit $m\to0$. 
The $16^3$ $g^{-2}=0.6$ point suggests that volume effects are small in this
regime.
This is similar to the findings of simulations with $N=2$ using the HMC
algorithm~\cite{Hands:2016foa}; however in that study there was no attempt to
take the $L_s\to\infty$ limit. Here we rectify that omission by plotting
extrapolated data from HMC simulations with $L_s=8,\ldots,40$ at the strongest
available coupling $g^{-2}=0.3$. Fortunately the conclusions of
\cite{Hands:2016foa} remain unchanged; there is no evidence for spontaneous
symmetry breaking, implying $N_c<2$. The contrast with the results
of Fig.~\ref{fig:Lsinfty} is particularly striking; whilst the condensate in the
quenched model shows no significant $m$-dependence, here the linear behaviour is
precisely that expected of a unitary field theory in its symmetric phase.

For $N=1$ at the strongest coupling examined $g^{-2}=0.3$,
$\langle\bar\psi\psi(m)\rangle$ is a factor of two or greater than data from the next
strongest coupling, and a linear extrapolation
$\lim_{m\to0}\langle\bar\psi\psi(m)\rangle=\Sigma\approx O(0.1)\not=0$ 
looks 
reasonable, particularly if the $16^3$ point is used at $m=0.01$.
This would be consistent with the spontaneous breakdown of U(2) symmetry due to
bilinear condensation at this
coupling, although non-linear extrapolations to a symmetric limit
$\langle\bar\psi\psi\rangle=0$ cannot at this stage be excluded.
If symmetry is indeed broken, 
on general grounds significant finite volume corrections are expected in the mesoscopic
regime $m\Sigma V\lapprox1$, and the data support this; note that
the dimensionless combination $m\Sigma V\approx1.5$ for the $12^3$, $m=0.01$ point.

In summary, Fig.~\ref{fig:N_1} presents strong evidence for the Thirring model
with $g^{-2}=0.3$ to exhibit  qualitatively very different behaviour from that
observed at weaker couplings, due to a  significant enhancement of fermion --
antifermion pairing. Finite-$L_s$ corrections are also much more important in
this regime, as illustrated in Fig.~\ref{fig:comparison}, and an $L_s\to\infty$ 
extrapolation proves key to interpreting the data. The simplest explanation is
that U(2) symmetry is spontaneously broken at the strongest coupling examined,
implying $N_c>1$.

\subsection{The Approach to $L_s\to\infty$}
\label{sec:Lstoinfty}

\begin{figure}[H]
    \centering
    \includegraphics[width=12.5cm]{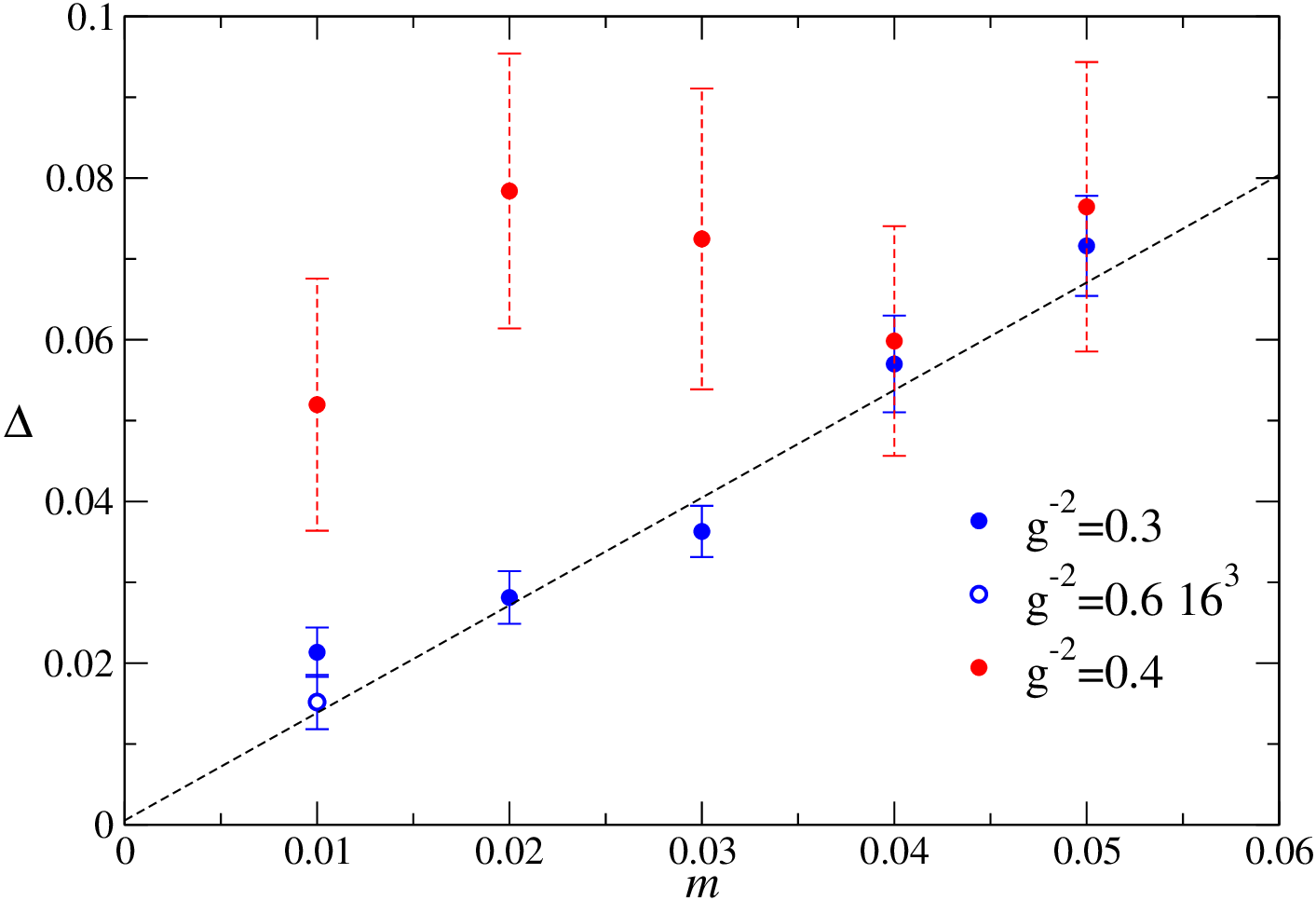}
\caption{(color online) The decay constant $\Delta(m)$ obtained from fits to (\ref{eq:expfit}) for 
$g^{-2}=0.3,0.4$.}
\label{fig:Deltaf}
\end{figure}

It is interesting to compare the $m$-dependence of the decay constant $\Delta$,
implicitly defined in (\ref{eq:expfit}), between different couplings. Of course, for a fixed
window in $L_s$, $\Delta$ is easier to pin down for data with large
curvature, corresponding to strong couplings and larger masses. For this
reason the large uncertainties on $\Delta$ from the weaker couplings $g^{-2}=0.5,0.6$
don't yield much of use; however results from the stronger couplings
$g^{-2}=0.3, 0.4$ plotted in Fig.~\ref{fig:Deltaf} show a marked 
contrast. Within sizeable uncertainties $\Delta(g^{-2}=0.4)\approx0.06$--$0.07$ is
approximately $m$-independent, whereas $\Delta(g^{-2}=0.3)\propto m$, the
linearity becoming more convincing still if the $16^3$ value is taken at
$m=0.01$. The straight line fit shown yields a slope 1.33(15), with intercept
consistent with passing through the origin. 
This is another hint of a qualitative difference in the behaviour of
the model at these two couplings. 

\begin{figure}[H]
    \centering
    \includegraphics[width=12.5cm]{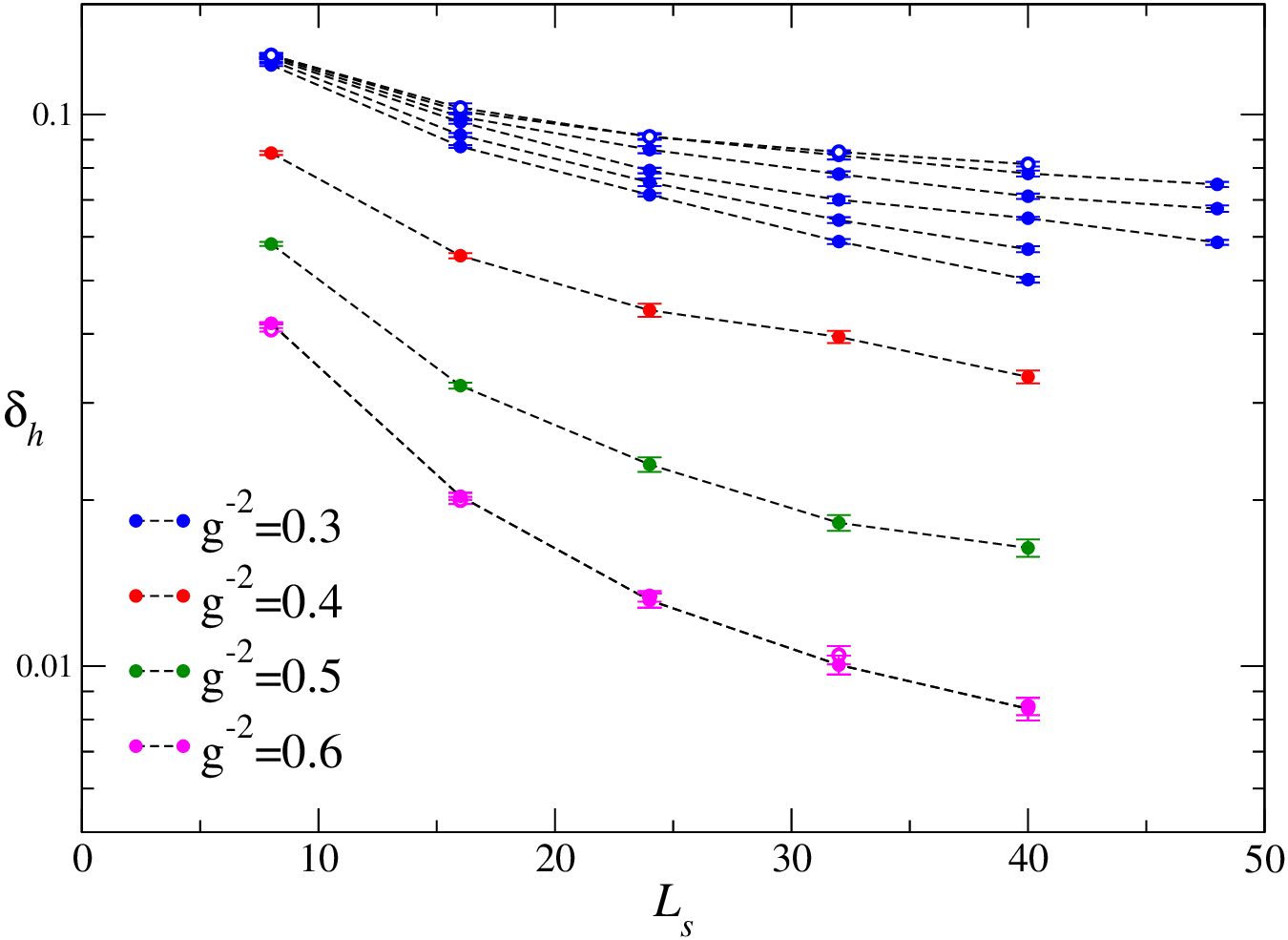}
\caption{(color online) The residual $\delta_h(L_s)$, defined in (\ref{eq:deltah}), for 
various $g^{-2}$ and $m=0.01$. For $g^{-2}=0.3$ results for all 5 mass values are shown,
along with $16^3$, $m=0.01$ (open circles).
For $g^{-2}=0.6$ results for $m=0.05$ are shown as open circles}
\label{fig:deltah}
\end{figure}

Another measure for the approach to the $L_s\to\infty$ limit is the
residual $\delta_h$ defined in (\ref{eq:deltah}). As shown in
\cite{Hands:2015qha}, it quantifies the difference between 
the U(2)-equivalent condensates $\langle\bar\psi\psi\rangle$ and the measured 
$i\langle\bar\psi\gamma_3\psi\rangle$,
and  should therefore vanish in a simulation respecting U(2) symmetry.
Results for $\delta_h(L_s)$ for various couplings are shown on a log scale in
Fig.~\ref{fig:deltah}. Just as in quenched QED$_3$ (see Fig.~2 of
\cite{Hands:2015qha}), $\delta_h$ is strongly
coupling-dependent. In all
cases the data is consistent with an asymptotic behaviour $\delta_h\propto e^{-cL_s}$
implying U(2) restoration in the large-$L_s$ limit; however the restoration
becomes slower as coupling increases. There is a marked difference between
$g^{-2}=0.6$, where $\delta_h$ is roughly $m$-independent, and
$g^{-2}=0.3$ where data from all 5 masses are plotted, and $c$ found apparently to
decrease systematically with $m$. At this strong coupling for $m=0.01$ $\delta_h$ is of the
same order of magnitude as the signal $i\langle\bar\psi\gamma_3\psi\rangle$ even for
$L_s=48$. For the larger $16^3$ lattice, $c$ is
smaller still; a similar trend was observed in \cite{Hands:2015qha}.

The findings of both Figs.~\ref{fig:Deltaf},\ref{fig:deltah} are consistent with
the extrapolation $L_s\to\infty$ used to obtain Fig.~\ref{fig:N_1}, and moreover
both display qualitative differences between strong and weak coupling, thus
supporting the argument that $g^{-2}=0.3$ and $g^{-2}=0.6$ lie in different
phases. However the approach to the large-$L_s$ limit becomes very slow in the symmetry broken phase in
the limit $m\to0$, which will almost certainly present practical difficulties in
future more refined simulations, and may also raise more conceptual problems
related to the existence of a U(2)-symmetric limit at strong coupling. 

\subsection{Meson Correlators}
\label{sec:mesons}

Finally we consider correlators of states formed from a fermion and an
antifermion, which by analogy with QCD will be referred to as mesons. 
We will focus on the sector with angular momentum $J=0$,
in which case four interpolating operators can be written. With a
choice of mass term $S_3$, they split into two scalars ($\bar\psi\psi$,
$\bar\psi\gamma_3\psi$) and two pseudoscalars ($\bar\psi\gamma_5\psi$,
$\bar\psi\gamma_3\gamma_5\psi$). In the event that a symmetry-breaking
condensate $i\langle\bar\psi\gamma_3\psi\rangle\not=0$ forms, then a 
Goldstone boson of either parity, interpolated by 
$\bar\psi\gamma_5\psi$ ($0^-$) and $\bar\psi\psi$ ($0^+$), is expected.
The other two states remain massive.

The DWF formulation was set out in \cite{Hands:2015qha}
in terms of ``primitive'' propagators
\begin{eqnarray}
C^{--}(x)&=&\mbox{tr}\left[S(m_3;0,1;x,L_s)P_-S^\dagger(m_3;0,1;x,L_s)P_-\right];\\
C^{+-}(x)&=&\mbox{tr}\left[S(m_3;0,1;x,1)P_+S^\dagger(m_3;0,1;x,1)P_-\right],
\end{eqnarray}
where the 2+1+1$d$ fermion propagator
$S(m_a;x,s;y,s^\prime)=\langle\Psi(x,s)\bar\Psi(y,s^\prime)\rangle_{m_a}$.
By construction $C^{\pm-}$ are real and positive.
It can be shown that the meson interpolated by $\bar\psi\gamma_5\psi$ has 
a propagator $C^{+-}+C^{--}$, while that interpolated by
$\bar\psi\gamma_3\gamma_5\psi$ has propagator $C^{+-}-C^{--}$. The
other two mesons in principle require additional fermion propagator calculations
with the flip $m_3\mapsto-m_3$; however in the context of quenched QED$_3$ it
was shown in \cite{Hands:2015qha} that the propagator interpolated by
$\bar\psi\psi$ becomes approximately equal to $\bar\psi\gamma_5\psi$, and
that of $\bar\psi\gamma_3\psi$ equal to $\bar\psi\gamma_3\gamma_5\psi$, in the limit
$L_s\to\infty$. Degeneracy of these opposite parity mesons is necessary for
U(2) symmetry restoration; we will not pursue this issue further, but rather 
confine our attention to the pseudoscalar channels interpolated by $\bar\psi\gamma_5\psi$ and
$\bar\psi\gamma_3\gamma_5\psi$, which will be referred to as Goldstone ($G$) and
non-Goldstone ($NG$) respectively.

\begin{figure}[H]
    \centering
    \includegraphics[width=12.5cm]{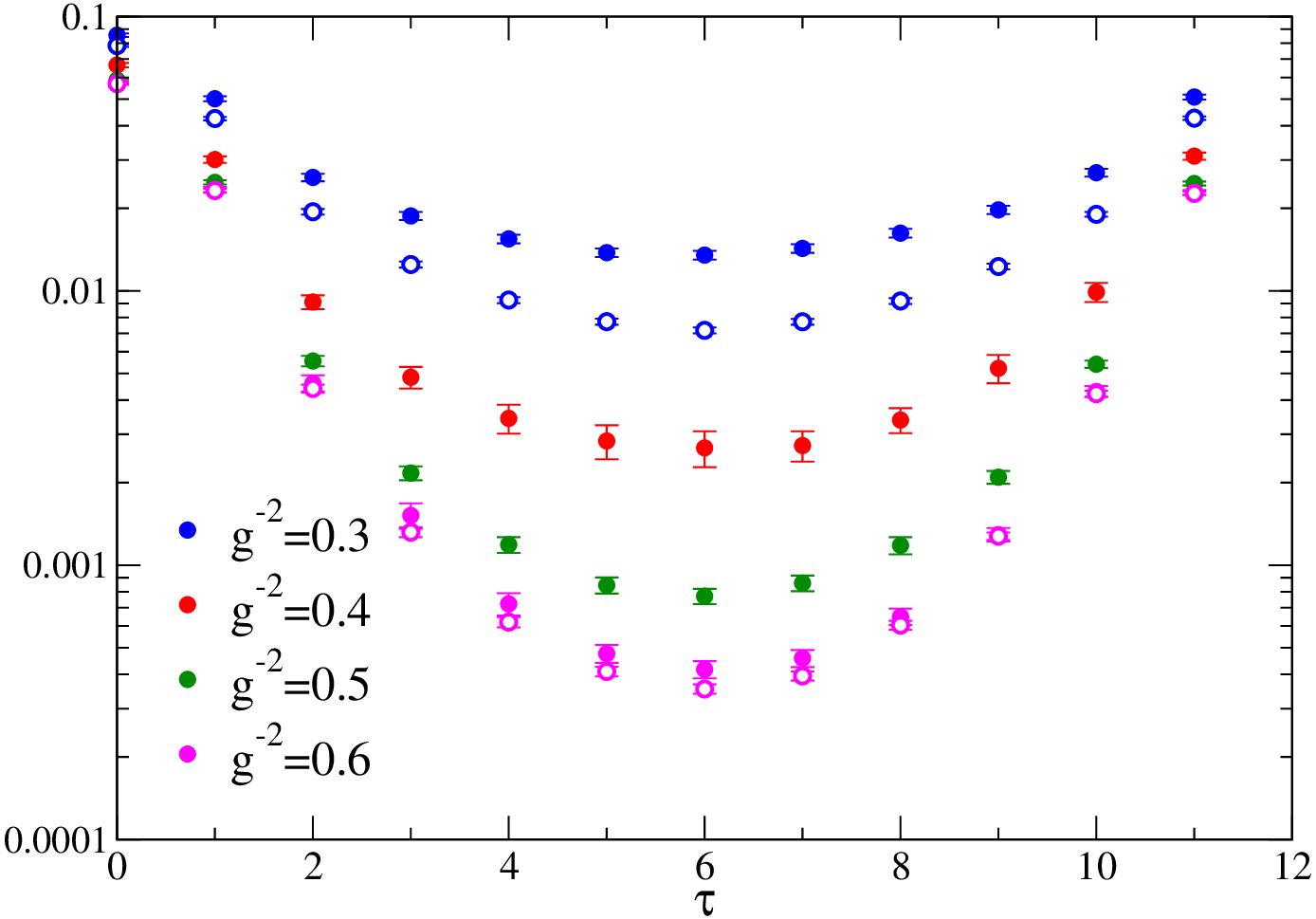}
\caption{(color online) Timeslice correlator for the Goldstone meson $C_G(\tau)$ 
on $12^3\times40$ for various $g^{-2}$ with
$m=0.01$ (filled symbols) and $m=0.05$ (open symbols).}
\label{fig:goldstone}
\end{figure}

\begin{figure}[H]
    \centering
    \includegraphics[width=12.5cm]{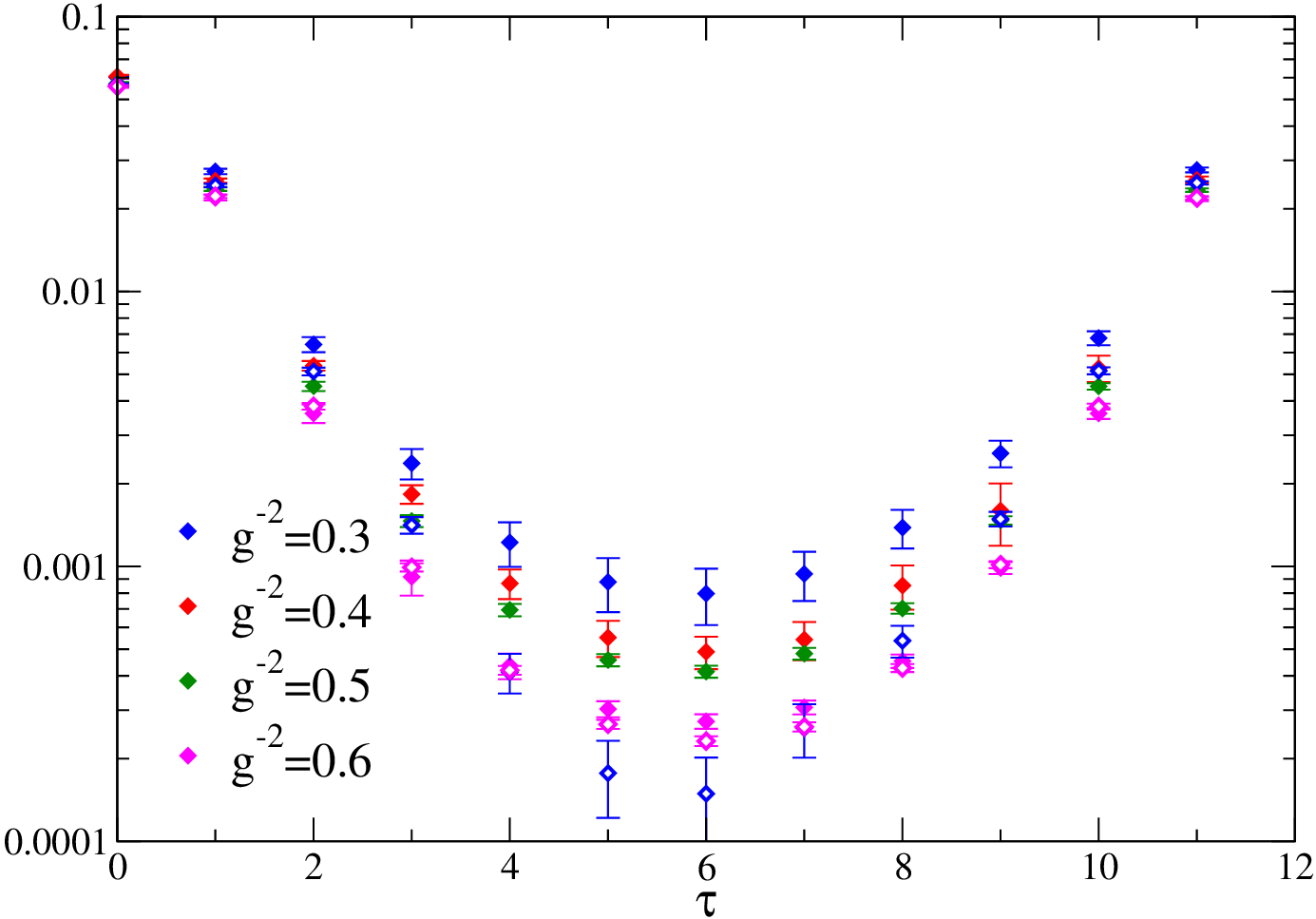}
\caption{(color online) Timeslice correlator for the non-Goldstone meson $C_{NG}(\tau)$ 
on $12^3\times40$ for various $g^{-2}$ with
$m=0.01$ (filled symbols) and $m=0.05$ (open symbols).}
\label{fig:nongoldstone}
\end{figure}

Meson timeslice correlators $C(\tau)=\sum_{\vec x}C(\vec x,\tau)$ were
calculated on a $12^3$ lattice with $L_s=40$ with $m=0.01$ at each 
coupling already investigated, with a minimum of 500 RHMC trajectories.
The primitive propagators were calculated every 5 trajectories by averaging over 5 point sources
located at random spacetime points, and the reconstructed $G$ and $NG$
correlators are plotted in Figs.~\ref{fig:goldstone} and
\ref{fig:nongoldstone} respectively.  Additional calculations with $m=0.05$
were performed at $g^{-2}=0.3$, 0.6.

A $12^3$ lattice is too far from both thermodynamic and zero-temperature
limits for any statements about the model's spectrum to be reliable; 
there is no sign of pure exponential decay corresponding to a simple
propagator pole, and as we shall see below it is also not safe at this stage to
infer anything regarding the residue.
Accordingly we
restrict ourselves to qualitative comments. In the $G$ channel
(Fig.~\ref{fig:goldstone}) there is a huge variation in signal size
as the coupling increases, with in particular a factor of 18 increase at the midpoint
$\tau=6$ between $g^{-2}=0.5$ and 0.3, and one of 5 between $g^{-2}=0.4$ and 0.3. 
Moreover, the impact of changing $m$ from
0.01 to 0.05 is far more pronounced at $g^{-2}=0.3$, where $C_G(6)$ decreases by
almost a half, and $g^{-2}=0.6$, where the decrease is less than 20\%. In the
$NG$  channel (Fig.~\ref{fig:nongoldstone}) the variation at the
midpoint between $g^{-2}=0.3,0.6$ is an order of magnitude smaller. 


Quantitively, the ratio $C_G(6):C_{NG}(6)$ increases from $\sim1.5$ at $g^{-2}=0.6$ to
$\sim17$ at the strongest coupling. In terms of primitive correlators, this implies 
that $C^{--}\ll C^{+-}$ at weak coupling, so that $G$ and $NG$ channels are
approximately degenerate,  but that $C^{--}\lapprox C^{+-}$ by
$g^{-2}=0.3$. This development is also reflected by the relatively large
errorbars in $C_{NG}$ at this coupling.

In view of the results for the would-be order parameter
$\langle\bar\psi\psi\rangle$ of Sec.~\ref{sec:condensate}, a natural
interpretation of these results is that U(2) symmetry spontaneously breaks
somewhere in the range $g^{-2}\in(0.5,0.3)$ and that both the increased magnitude of 
$C_G$ and its enhanced sensitivity to a change in $m$ is due to its developing
into a true Goldstone boson. 

It is interesting to compare these results with those presented in Fig.~15 of
Ref.~\cite{Hands:2016foa} for the would-be Goldstone meson on $12^2\times24$ in the surface
formulation of the Thirring model with $N=2$. In that case $C_G(\tau)$ manifests a 
$\tau$-independent plateau for $5\lapprox\tau\lapprox20$ over a range of couplings, 
interpreted in \cite{Hands:2016foa} as being due to fermion propagators
reconnecting only after one of them loops around the timelike extent of the system. In other
words, the surface formulation does not appear to support mesonic bound states.
With the caveats already discussed, the meson correlators of
Figs.~\ref{fig:goldstone},\ref{fig:nongoldstone} do appear to resemble those of
conventional mesons. This is the first hint that the bulk formulation is the
preferred approach to the Thirring model with DWF.

\begin{figure}[H]
    \centering
    \includegraphics[width=12.5cm]{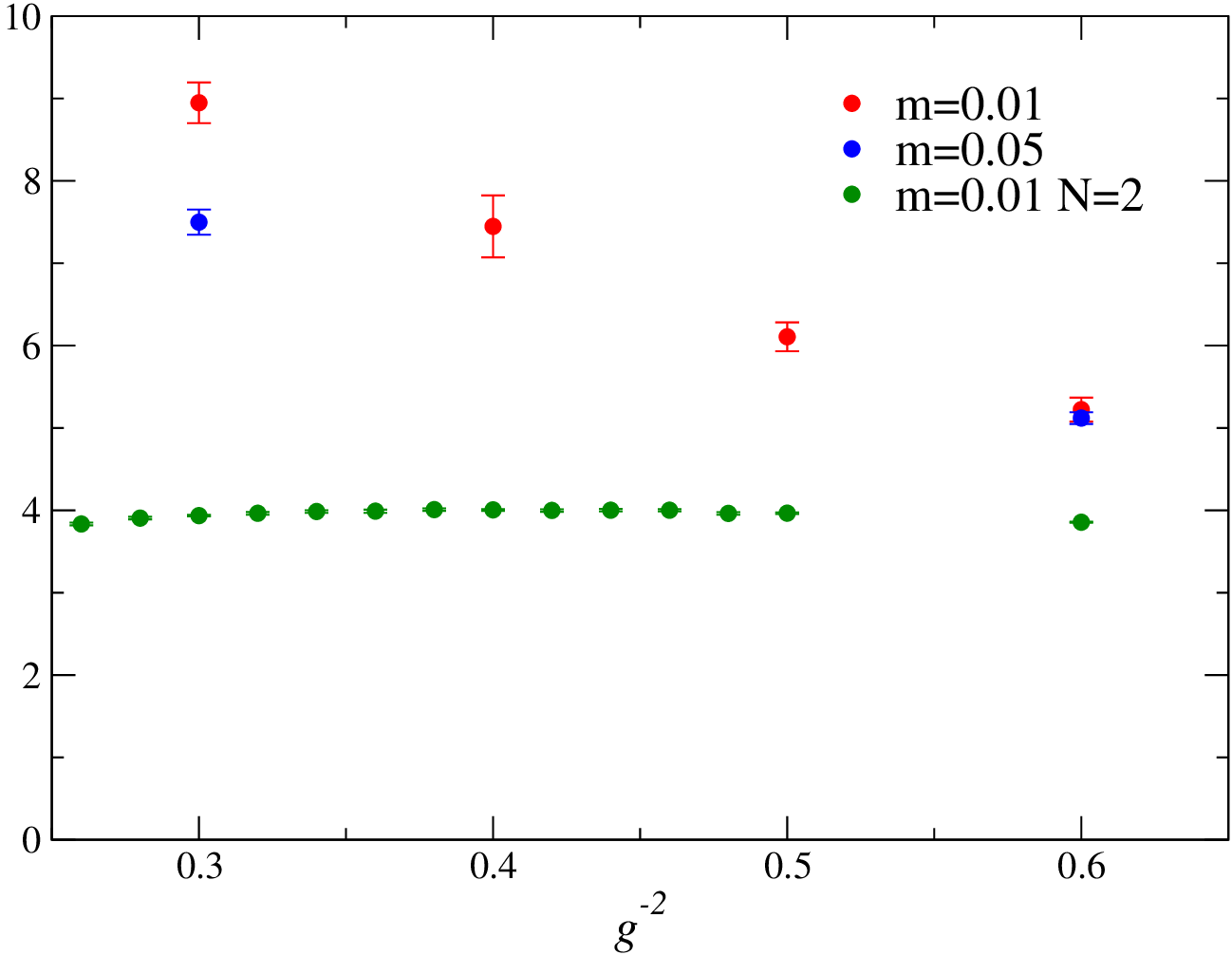}
\caption{(color online) The ratio $\langle\bar\psi\psi\rangle/m\chi_\pi$ vs. $g^{-2}$ on
$12^3\times40$ for various $m$, together with corresponding data taken at $N=2$
on $12^3\times16$ and $m=0.01$~\cite{Hands:2016foa}.}
\label{fig:ward}
\end{figure}
Finally, we consider more a formulational issue by examining the axial Ward
identity, first considered in this context in \cite{Hands:2016foa}. For a system
with the U(2) symmetry anticipated in the $L_s\to\infty$ limit, the following
identity relating the order parameter with the integrated correlator
holds\footnote{The factor of 2 after the second equality in (\ref{eq:ward}) reflects the
contributions of $C^{-+}, C^{++}$.}:
\begin{equation}
{{i\langle\bar\psi\gamma_3\psi\rangle}\over m}=\sum_x
\langle\bar\psi\gamma_5\psi(0)\bar\psi\gamma_5\psi(x)\rangle=2\sum_\tau C_G(\tau)\equiv\chi_\pi.
\label{eq:ward}
\end{equation}
The ratio $\langle\bar\psi\psi\rangle/m\chi_\pi$ is plotted vs. $g^{-2}$ in
Fig.~\ref{fig:ward}, together with bulk formulation results for $N=2$ on $12^3\times16$ 
~\cite{Hands:2016foa}. The clear issues are that the ratio is neither constant,
nor equal to unity, as required by (\ref{eq:ward}). Ref.~\cite{Hands:2016foa}
suggested this is due to a non-trivial relation between either or both of the
fermion mass $m$  and the physical fields $\psi\bar\psi$  defined in
(\ref{eq:4to3}), and their continuum counterparts (amusingly, on the vertical scale used
in Fig.~\ref{fig:ward} the data provoking this speculation now looks rather
constant as a function of $g^{-2}$). This would mean that bilinear operators
and/or the fermion mass $m$ would need
to be renormalised for the Ward identity to apply. Fig.~\ref{fig:ward} suggests
these considerations become still more important at strong coupling for $N=1$;
indeed, at $g^{-2}=0.3$ even the effect of changing $m$ results in a marked
renormalisation. Recall the ratio was observed to be $m$-independent for
$N=2$~\cite{Hands:2016foa}. Strong renormalisations depending on both $g^2$ and
$m$ cannot be ruled out for the parameter regime studied in this paper; moreover
we draw some encouragement from the hints in Fig.~\ref{fig:ward} that the effect
is smooth as $g^{-2}$ ranges from 0.6 to 0.3, consistent with UV physics, and in
contrast with the sharp changes over the same range reported in the rest of
this section, associated with a symmetry breaking phase transition.

\section{Discussion}
\label{sec:discussion}

The main result of this paper is that the spontaneous breakdown of the U($2N$)
symmetry present for massless reducible fermions in 2+1$d$ can be
demonstrated in simulations of an interacting field theory using domain wall fermions. The proof of concept
was given in the quenched limit in Sec.~\ref{sec:quenched}, where the
importance of taking the $L_s\to\infty$ limit {\em before\/} the $m\to0$ limit
was shown. Next, simulations of the unitary $N=1$ model with a newly-developed RHMC
algorithm, discussed in Sec.~\ref{sec:N=1},
yielded results following the same procedure consistent with unbroken U(2)
symmetry for $g^{-2}\geq0.4$, but with enhanced bilinear condensation at the
strongest available coupling $g^{-2}=0.3$, consistent with a non-vanishing
intercept in the $m\to0$ limit signalling the breaking of U(2) (see
Fig.~\ref{fig:N_1}). Meson
correlators on admittedly small spacetime volumes were consistent with the
Goldstone spectrum expected for the breaking pattern U(2)$\to$U(1)$\otimes$U(1)
(see Figs.~\ref{fig:goldstone},\ref{fig:nongoldstone}).
The most natural conclusion is that there is a symmetry-breaking phase
transition at some $g_c^{-2}\in(0.3,0.4)$, that the critical flavor number in
the 2+1$d$ Thirring model satisfies $1<N_c<2$, and that there is the potential
for a QCP in the $N=1$ model described by a strongly-interacting local unitary
quantum field theory. 
Final confirmation of this important result must await 
further simulations permitting enhanced control over both $V\to\infty$ and
$m\to0$ limits; until then strictly the bound we have found is $0<N_c<2$.
The large disparity with the staggered Thirring model
result $N_c=6.6(1)$~\cite{Christofi:2007ye} is a dramatic 
indicator of the importance of the faithful rendition of global symmetries when
modelling 
strongly-interacting systems.

As a bonus, the form of the meson correlators 
strongly suggests the preferred
formulation of the Thirring model with DWF uses the bulk formulation of the
vector auxiliary, clearing up an outstanding issue from previous
work~\cite{Hands:2016foa}. 
However the quenched results  of Sec.~\ref{sec:quenched} revealing symmetry
breaking in the surface model at very strong coupling mandate further
investigation of this formulation. The question of the most natural 
formulation of this (or any strongly-interacting) model permitting
systematic numerical investigation remains open; here it is prudent to recall
that simulations of the Thirring model with SLAC fermions find
$N_c<1$~\cite{Wellegehausen:2017goy}.

The $N=1$ bulk simulations also raise some concerns. The decay constant $\Delta$
governing the $L_s\to\infty$ extrapolation seems to follow $\Delta\propto m$
for $g^{-2}<g_c^{-2}$
(see Fig.~\ref{fig:Deltaf}), implying that there may be both practical and even
conceptual difficulties reaching the massless limit in the broken phase.
The residual $\delta_h$ parametrising the explicit U(2$N$) symmetry
breaking at finite $L_s$ is also rather large and slowly-decaying in this
regime (see Fig.~\ref{fig:deltah}). This suggests rather careful attention will
need to
be paid to the question of U(2$N$) symmetry in future studies of the
broken phase. We also remark that a further outstanding issue is the locality of
the associated 2+1$d$ overlap operator, which governs the validity of the
U($2N$) in terms of global symmetry rotations on local fields in
2+1$d$~\cite{Hands:2015dyp,Luscher:1998pqa}. An understanding of each of these issues 
is a precondition for a satisfactory operational definition of quantum field
theories of strongly-interacting fermions.

In future work we plan to implement simulation code with improved
performance to counter the considerable numerical effort required for the inversion of
${\cal M}^\dagger{\cal M}$ at the strong couplings relevant for symmetry
breaking. The need for further improvements in the
invertor algorthm, and even in the DWF formulation following the ideas of
\cite{Chiu:2002ir}, should not be ruled out. The next step is a more refined scan
of the $N=1$ theory in the critical region $g^{-2}\in(0.3,0.4)$ with the goal
of first locating and then characterising the critical point at $g_c^{-2}$.
The potential difficulty of correctly capturing critical fluctuations is 
highlighted in the inset of Fig.~\ref{fig:comparison}.
Finally, using the control over global symmetries furnished  by DWF it will be
straightforward to examine the effect of a U(2$N$) and parity-invariant 
``Haldane'' contact interaction $(\bar\psi\gamma_3\gamma_5\psi)^2$, which in
Ref.~\cite{Gies:2010st} was found to be a component of the interaction at the
fixed point. Exploratory results in this direction were reported in
\cite{Schmidt:2016rtz}.

\section*{Acknowledgements}
This work was supported 
in part by STFC grant ST/L000369/1. Numerical work was
performed on a PC cluster funded by Swansea University's College of Science. 
I have been greatly helped by discussions with Ed Bennett, Nikhil Karthik,  
Tony Kennedy, Michele Mesiti, Andreas Wipf and Jude Worthy.

\appendix
\section{The determinant in the bulk formulation}
\label{app:A}
Write the fermion action in (\ref{eq:action}) as 
\begin{equation}
\bar\Psi{\cal M}\Psi\equiv\bar\Psi D_W\Psi+\bar\Psi D_3\Psi+m_aS_a
\end{equation}
with
\begin{equation}
D_W\equiv\gamma_\mu D_\mu-(\hat
D^2+M);\;\;\;D_3\equiv\gamma_3\partial_3-\hat\partial_3^2.
\end{equation}
The symbol $\partial$ is reserved for operators with no
dependence on the auxiliary $A_\mu$.
These definitions yield the properties $\hat D^2=\hat D^{2\dagger}\not=\sum_\mu
D_\mu^2$ and
$\hat\partial_3^2=\hat\partial_3^{2\dagger}\not=\partial_3\partial_3$.
We first note that for $a=h,3$ the identity
$\gamma_5{\cal M}\gamma_5={\cal M}^\dagger$~\cite{Hands:2015qha}
ensures that $\mbox{det}{\cal M}^2$ is positive and hence $\mbox{det}{\cal M}$
real.

In the Dirac basis $\gamma_\mu=\sigma_{\mu+1}\otimes\tau_3$ ($\mu=0,1,2$) and
$\gamma_3=\One\otimes\tau_2$, where $\vec\sigma$ and $\vec\tau$ are Pauli
matrices, and setting $m_a=0$, we find
\begin{eqnarray}
{\cal M}&=&\left(\begin{matrix}
{\sigma_{\mu+1}D_\mu-(\hat D^2+M+\hat\partial_3^2) &  -i\partial_3\cr
i\partial_3 & -\sigma_{\mu+1}D_\mu-(\hat D^2+M+\hat\partial_3^2)\cr}
\end{matrix}\right)\nonumber\\
&\equiv&\left(\begin{matrix}
{D_{\hat W}-\hat\partial_3^2 &  -i\partial_3\cr
i\partial_3 & D_{\hat W}^\dagger-\hat\partial_3^2\cr}
\end{matrix}\right)
\end{eqnarray}
so that
\begin{equation}
\mbox{det}{\cal M}=\mbox{det}(-i\partial_3)\mbox{det}
[i\partial_3-(D_{\hat W}-\hat\partial_3^2)^\dagger(-i\partial_3)^{-1}
(D_{\hat W}-\hat\partial_3^2)].\label{eq:detD}
\end{equation}
Now, if the commutator $[\partial_3,D_{\hat W}-\hat\partial_3^2]=0$, then
(\ref{eq:detD}) can be rearranged to read
\begin{equation}
\mbox{det}{\cal M}
=\mbox{det}[\partial_3^\dagger\partial_3+(D_{\hat W}-\hat\partial_3^2)^\dagger
(D_{\hat W}-\hat\partial_3^2)]
=\mbox{det}(B^\dagger B)\mbox{det}[C^\dagger C+\One]
\label{eq:perfect}
\end{equation}
where $B=\partial_3$, $C=(D_{\hat W}-\hat\partial_3^2)\partial_3^{-1}$, and the last step
follows if $B$ is invertible. Then  $\mbox{det}{\cal M}$ would be  positive definite,
and moreover ${\cal M}$ could be represented as a positive operator making it
possible to simulate using bosonic pseudofermions. 

Now let's examine the commutator. The contributions
$[\partial_3,D_\mu]=[\partial_3,\hat D^2+M]=0$, which follows provided the link
connections obey $U_{\mu,x}=U_{\mu,x\pm\hat3}$ and
$U_{\mu,x}^\dagger=U_{\mu,x\pm\hat3}^\dagger$. This is the case both for gauge
theories and for the bulk formulation of the Thirring model; in each case the
connection is ``3-static'', ie. $\partial_3 U_{\mu,x}=0$. However, the remaining
part is non-vanishing:
\begin{equation}
[\partial_3,\hat\partial_3^2]={1\over2}\delta_{x,y}\left(\delta_{x_3,1}-\delta_{x_3,L_s}\right).
\end{equation}
Whilst a simple physical interpretation of this term is obscure, it is
clear the obstruction to proving the positivity of $\mbox{det}{\cal M}$ has its
origin in the open boundary conditions imposed at the walls. Now consider a
Dirac basis $\gamma_\mu=\sigma_{\mu+1}\otimes\tau_2$,
$\gamma_3=\One\otimes\tau_3$ so that
\begin{equation}
{\cal M}=\left(\begin{matrix}
{\partial_3-(\hat D^2+M+\hat\partial_3^2)&  -i\sigma_{\mu+1}D_\mu\cr
i\sigma_{\mu+1}D_\mu & -\partial_3-(\hat D^2+M+\hat\partial_3^2)\cr}
\end{matrix}\right).
\end{equation}
In this case the obstruction to proving positivity turns out to be the 
non-vanishing commutator
\begin{eqnarray}
[D_\mu,\hat D^2]=&-{1\over4}&
\sum_\nu\Bigl[
(U_{\mu x}U_{\nu x+\hat\mu}-U_{\nu x}U_{\mu
x+\hat\nu})\delta_{x+\hat\mu+\hat\nu,y}\cr
&+&(U_{\mu x}U^\dagger_{\nu x+\hat\mu-\hat\nu}-U^\dagger_{\nu x-\hat\nu}U_{\mu
x-\hat\nu})\delta_{x+\hat\mu-\hat\nu,y}\cr
&+&(U_{\nu x}U^\dagger_{\mu x-\hat\mu+\hat\nu}-U^\dagger_{\mu x-\hat\mu}U_{\nu
x-\hat\mu})\delta_{x-\hat\mu+\hat\nu,y}\cr
&+&(U^\dagger_{\nu x-\hat\nu}U^\dagger_{\mu x-\hat\mu-\hat\nu}-U^\dagger_{\mu
x-\hat\mu}U^\dagger_{\nu
x-\hat\mu-\hat\nu})\delta_{x-\hat\mu-\hat\nu,y}
\Bigr],\label{eq:comm}
\end{eqnarray}
which by construction is evenly distributed throughout the bulk.
The commutator (\ref{eq:comm}) vanishes for configurations in which both the
``plaquette'' $U_{\mu\nu}=1$ and $U_\mu^\dagger U_\mu=1$, which for the Thirring model
is expected to be reached only in the limit $g^2\to0$.

We conclude that $\mbox{det}{\cal M}$ is real but not in general positive,
motivating the use of the RHMC algorithm to simulate the functional measure 
$\mbox{det}({\cal M^\dagger M})^{1\over2}$ outlined in
Sec.~\ref{sec:formulation}. We note that there is no such obstruction 
for HMC simulations of $\mbox{det}{\cal M}$ using twisted-mass
Wilson fermions~\cite{Karthik:2015sgq} or overlap
fermions~\cite{Karthik:2016ppr}, both of which have been recently used to study
QED$_3$.

\end{document}